# Laterally Differentiated Polymorphs: a route to multifunctional nanostructures

Pete E Lauer[1]*, Kensuke Hayashi[1,2,3]*, Yuichiro Kunai[4,1], Ondřej Wojewoda[5], Jan Klíma[5], Ekaterina Pribytova[5], Michal Urbánek[5], Aubrey Penn[6], Takayuki Kikuchi[3], Renzhi Ma[3], Takayoshi Sasaki[3], Takaaki Taniguchi[3], Caroline A. Ross[1+]

[1]Department of Materials Science and Engineering, Massachusetts Institute of Technology, Cambridge, Massachusetts 02139, USA

[2]Department of Materials Physics, Nagoya University, Furo-cho, Chikusa-ku, Nagoya, Japan, 464-8603

[3]Research Center for Materials Nanoarchitectonics (MANA), National Institute for Materials Science (NIMS), 1-1 Namiki, Tsukuba, Ibaraki 305-0044, Japan

[4]Advanced Research Core, Fujikura Limited, 1-5-1, Kiba, Kouto-ku, Tokyo 135-8512, Japan

[5]CEITEC BUT, Brno University of Technology, Purkyňova 123, 621 00 Brno, Czech Republic

[6]MIT.nano, Massachusetts Institute of Technology, Cambridge, Massachusetts 02139, USA

*equal contributions

+ caross@mit.edu

## Abstract

Multifunctional materials can exhibit emergent behavior from the coupling of two or more different properties. For example, coupling between magnetic and ferroelectric order enables electrical control of the magnetic state, enabling for example magnetoelectric memory or logic devices that combine the nonvolatility of magnetic order with the energy efficiency of voltage control. Magnetic iron garnets have outstanding magnonic and magnetooptical properties making them valuable in a range of technologies, but they have not been successfully incorporated into thin film two-phase magnetoelectric nanocomposites. Taking advantage of heterogeneously patterned substrates, this work demonstrates the engineering of garnet-perovskite composites in which both phases are polymorphs with the same composition but dramatically different structures and properties. Applying an electric field to the perovskite phase modulates the magnon dispersion and magnetooptical response of the garnet, opening a path to voltage-controlled garnet devices.

Multifunctional materials with cross-coupled properties provide a platform for a wide array of next-generation devices. Magnetoelectrics are an important class of multifunctional materials, exhibiting coupling between ferroelectric (FE) and ferro- or ferrimagnetic (FM) order parameters which enables electric-field modulation of magnetization or magnetic-field modulation of polarization. Such cross-coupling, allowing for convenient voltage-driven manipulation or switching of the magnetic order parameter, could enable numerous low-power, high-speed applications in magnetic memory[1–3], spintronics[4], and magneto-optical devices[5]. Due to the scarcity[6] and low magnetoelectric coupling strength of single-phase multiferroic materials at room temperature, two-phase multiferroics have attracted considerable attention. These FM/FE composites can exhibit magnetoelectricity when the magnetostrictive and piezoelectric effects of the two components are coupled through interfacial strain transfer. For example, a piezoresponse in the FE component excited by an applied electric field causes a strain in the FM component which drives a change in its magnetic anisotropy[7].

Vertically aligned nanocomposites (VANs) provide an advantageous architecture for two-phase thin film multiferroics as their large interfacial area perpendicular to the substrate facilitates strain transfer and enhances magnetoelectric coupling[8–11]. VANs are typically grown via codeposition of two mutually insoluble epitaxial phases using pulsed laser deposition (PLD) or sputtering. The most commonly studied multiferroic VANs consist of a FM spinel (e.g. $CoFe_2O_4$, CFO[12–16]) and a FE perovskite (e.g. $BaTiO_3$, BTO[12–14] or $BiFeO_3$, BFO[15,16]) grown on a perovskite substrate such as $SrTiO_3$ (STO), in which FM pillars or fins are embedded in the FE matrix. The size, geometry, and spacing of the self-assembled phases depends on substrate orientation, volume fraction of phases, and growth parameters, resulting in pillar diameters from 10s – 100s nm in size. Templated nucleation and growth of the phases has been accomplished by patterning seeds of one component by liftoff or etching[17,18], by modifying the substrate surface termination[11], by etching pits in the substrate through a lithographic mask[19], or by patterning using a focused ion beam[20–22]. These VAN architectures have been proposed for memories, sensors, memristors and other devices[23,24].

Iron garnets, exemplified by yttrium iron garnet ($Y_3Fe_5O_{12}$, YIG), have ultra-low Gilbert damping, high resistivity, and magnetooptical activity, making them highly desirable for applications in RF oscillators/filters, magnonic devices, and photonic integrated circuits[25–27]. Garnet/perovskite VANs would therefore offer transformative opportunities for designing magnetoelectric devices that incorporate garnets with electrically-tunable properties. However, the integration of garnets into VANs has been challenging[28–30]. First, epitaxial growth is prevented by the structural incompatibility between garnets (lattice parameter $a \cong 1.2$ nm) and perovskites ($a \cong 0.4$ nm). Second, in spinel/perovskite VANs, phase separation is promoted because the large cations (Bi, Y, rare earths, Ba, etc.) are insoluble in the spinel phase, but this mechanism is not available in a garnet/perovskite composite in which the large cations can be accommodated in both phases. As a result, there has been very limited success in synthesizing garnet/perovskite VANs. Co-deposition of YIG and BTO on a $Gd_3Ga_5O_{12}$ (GGG) garnet substrate yielded a YIG film with defective regions containing Ba and Ti, rather than a two-phase VAN[30], whereas codeposition on a perovskite substrate yielded BTO pillars embedded in polycrystalline YIG of nonuniform thickness[28,29], and with secondary phases such as $YFeO_3$ (YFO) and $BaFe_{12}O_{19}$ present after annealing[29]. Garnet films have also been transferred or deposited onto piezoelectric substrates to make macroscopic magnetoelectric bimorphs[31–33]. However, in these structures strain transfer takes place along an in-plane interface, the deposited garnet is polycrystalline, and there is no capability for local control of patterned garnet features such as waveguides by a surrounding piezoelectric phase.

Here we present a methodology to create multifunctional composites consisting of a FM garnet and a FE perovskite with the same composition, designated as *laterally differentiated polymorphs* (LDPs). Our LDPs are created by pre-patterning regions of a thin perovskite seed layer or perovskite-structured nanosheet onto a garnet substrate and subsequently depositing a film of a composition such as $(Bi+Y)_zFe_yO_{1.5(z+y)}$. Templated by the heterogeneous substrate, perovskite/garnet LDPs are created for all z:y ratios between 3:5 (the ideal garnet stoichiometry) and 1:1 (the ideal perovskite stoichiometry). Thus, the LDP consists of garnet and perovskite polymorphic phases with lateral feature sizes of 50 nm and above and the *same* composition, but dramatically *different* structures and properties. We demonstrate magnetoelectric coupling in LDPs with compositions $Bi_{2.25}Y_{0.75}Fe_5O_{12}$ and $Bi_3Fe_5O_{12}$ in which an electric field applied to the perovskite phase modulates the magnon modes and the magnetooptical hysteresis loop of the garnet. These magnetoelectric composites unleash the potential of garnets for voltage-controlled spintronic, magnonic, and photonic devices.

**Synthesis of Laterally Differentiated Polymorphs**

Our fabrication of garnet-perovskite LDPs is based on the power of epitaxy to stabilize thin film crystal structures with non-equilibrium compositions. Taking $Y_zFe_yO_{1.5(z+y)}$ as an example, the equilibrium phase is YIG for z:y = 3:5 and the perovskite-derived orthoferrite $YFeO_3$ (YFO) for z:y = 1:1, with intermediate compositions corresponding to a two-phase mixture. However, films grown with z:y = 1:1 on GGG substrates yield single-phase films of Fe-deficient YIG, and films grown with z:y = 3:5 on STO substrates yield Y-deficient perovskite-derived YFO films[34,35]. Therefore, epitaxy with the substrate, rather than the cation ratio, determines which phase forms. To synthesize LDPs, we therefore developed a substrate with distinct surface regions that stabilize either garnet or perovskite-derived structures.

Heterogeneous substrates were formed by patterning a thin seed layer of a perovskite on a garnet substrate using electron beam lithography (Figure 1; Figure S1 in the Supplementary File). The seed layer was grown by PLD onto a resist pattern without heating the substrate, and after liftoff, the patterned seed layer was crystallized by a rapid thermal anneal (RTA) in oxygen. To prevent the seed layer from crystallizing as a garnet phase, we select Sr-containing perovskites (STO or $La_{0.7}Sr_{0.3}MnO_3$, LSMO) for the seed layer due to the very low solubility of Sr in garnet. The resulting substrate surface presents regions of polycrystalline perovskite and regions of single crystal garnet. A second strategy to create a heterogeneous substrate consists of dispersing flakes of 2-dimensional perovskite-structure calcium niobate nanosheets ($Ca_2Nb_3O_{10}$, CNO)[36] onto a GGG substrate, leading to (001)-oriented perovskite-compatible regions with ~55% coverage (Figure S2). Finally, films with composition $(Bi+Y)_zFe_yO_{1.5(z+y)}$ were deposited onto the heterogeneous substrates, forming epitaxial garnet on the exposed regions of the garnet substrate, and perovskite-derived structures, referred to as perovskites for brevity (orthoferrites for Y-Fe oxides and rhombohedral or tetragonal structures for $BiFeO_3$) above the STO, LSMO or CNO regions.

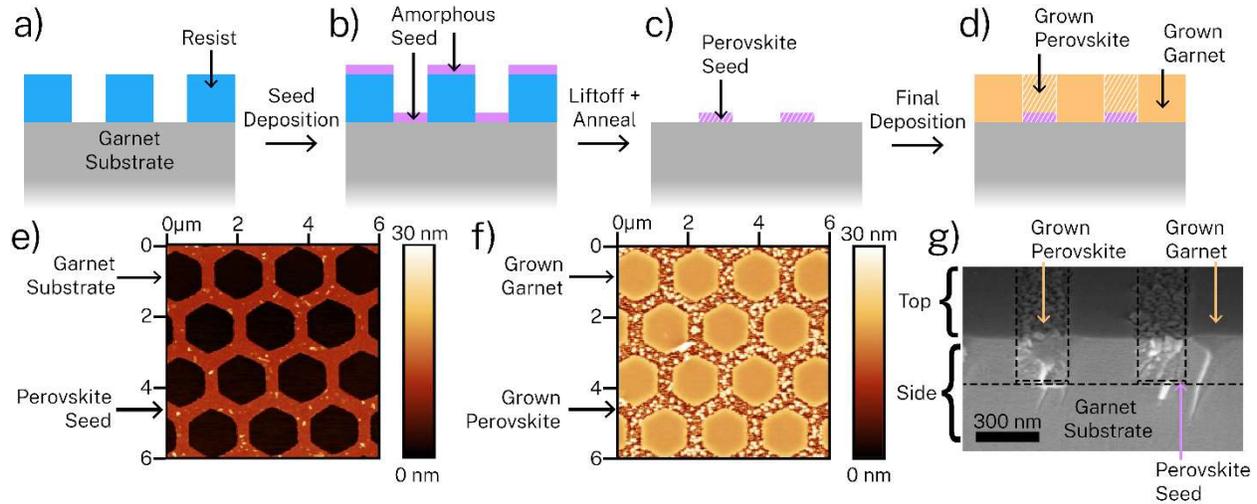

**Figure 1: Fabrication of Laterally Differentiated Polymorphs.** (**a-d**) Schematic representation of the process consisting of (**a**) spincoating, exposure, and development of the resist; (**b**) seed layer deposition without heating, (**c**) lift-off then annealing, and (**d**) final layer deposition, forming perovskite on the perovskite seed and garnet on the GGG surface. (**e**) AFM scan of a garnet substrate with 9.6 nm of patterned STO crystallized at 750 °C for 300 s. The polycrystalline STO seed has a root mean square surface roughness of $R_q$ = 1.47 nm compared to the exposed GGG regions with $R_q$ = 0.33 nm. (**f**) AFM scan of the surface of a 220 nm thick $Y_3Fe_5O_{12}$ LDP film grown on the template shown in (e). The polycrystalline perovskite regions have $R_q$ = 6.35 nm whereas the epitaxial single-crystal garnet regions have $R_q$ = 0.61 nm. (**g**) 67° tilted view of the cross section of a 220 nm thick $Y_3Fe_5O_{12}$ LDP displaying both the top surface (darker region), and the fractured cross-section surface (lighter region).

An example of the surface topography of a patterned perovskite seed (9.6 nm thick STO on (111) GGG) and the resulting LDP (220 nm thick, composition $Y_3Fe_5O_{12}$) is given in Figure 1ef, with Figure 1g showing a cross-section. Scanning transmission electron microscopy (STEM) and energy dispersive spectroscopy (EDS) reveal the coexisting garnet and perovskite phases and their compositions within a 382 nm thick LDP of composition $Y_3Fe_5O_{12}$ grown on a 12.2 nm thick STO seed, Figure 2. YIG grows epitaxially on the GGG whereas polycrystalline perovskite grows on the STO seed. Remarkably, the polycrystalline perovskite and single crystal garnet regions are polymorphs, having the same Y:Fe ratio despite their different crystal structures as evidenced by EDS, Figure 2b and Figure S6, i.e. the garnet is $Y_3Fe_5O_{12}$ and the perovskite is Fe-rich YFO, ($Y_{0.75}Fe_{0.25})FeO_3$. (The increased image brightness in the garnet phase of Figure 2a is a result of enhanced electron channeling through the single crystal garnet, not due to a stoichiometry difference.) High-angle annular dark field (HAADF) images, Figure 2c, confirm epitaxial growth of YIG on the GGG regions, while the STO seed layer and the Fe-rich YFO are polycrystalline. Tilting reveals zone axes of the STO and Fe-rich YFO, and a [100] zone axis of the Fe-rich YFO orthorhombic structure is visible in Figure 2d.

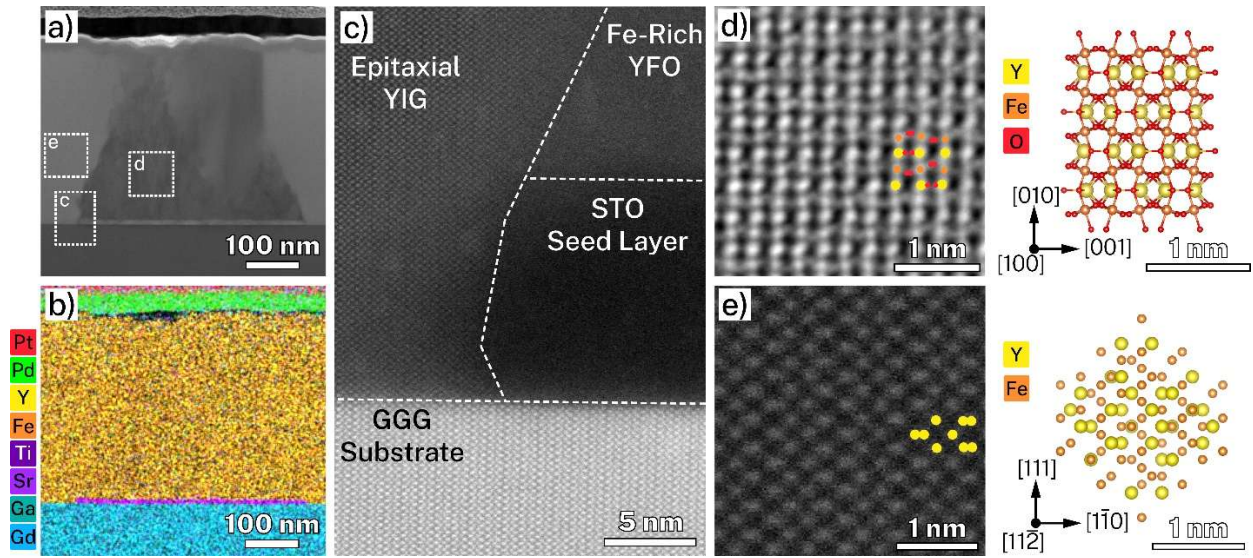

**Figure 2: Structure of an LDP with composition $Y_3Fe_5O_{12}$.** (**a**) STEM cross section of the 382 nm thick LDP showing the perovskite at center surrounded by garnet. (**b**) EDS scan of the same region shown in (a), showing that the garnet and perovskite are polymorphs with the same composition. (**c**) STEM cross section showing the substrate, single crystal YIG, and the polycrystalline STO seed and Fe-rich YFO perovskite. (**d**) STEM image taken with integrated differential phase contrast of Fe-rich YFO and the orthoferrite crystal structure along [100]; cation and oxygen columns are visible. (**e**) STEM image of YIG and schematic of the garnet crystal structure (with O atoms hidden) along [11$\bar{2}$].

Garnet films with cation ratios of Y:Fe = 3.5:4.5 and 1:1 were also synthesized on (111) GGG substrates (Figure S5ab). As the composition moves from Y:Fe = 3:5 to Y:Fe = 1:1, the lattice parameter of the garnet phase increases and the magnetization decreases, consistent with the incorporation of Y onto the octahedral sites to form $Y_{Fe}$ antisite defects which lower the Curie temperature leading to a lower room temperature magnetization[35]. Films grown on a STO-coated

GGG substrate instead formed perovskite structures across the composition range of Y:Fe = 3:5 to 1:1, while films grown on heterogeneously patterned substrates formed LDPs, for example consisting of YFO with Y-rich YIG for Y:Fe = 1:1, or Fe-rich YFO with Y-rich YIG for Y:Fe = 3.5:4.5.

These results show that across the entire composition range of Y:Fe = 1:1 to 3:5, the phase that forms (garnet or perovskite) is defined entirely by epitaxy and not by composition. Moreover, once the two phases are nucleated on a heterogeneous substrate, they continue to grow with vertical interfaces even for LDP thicknesses of 2 μm (Figure S4g).

In LDPs grown on GGG substrates with CNO nanosheets, the crystal orientation of the perovskite regions is templated by the CNO, leading to the pseudocubic (pc) [001] of the perovskite oriented out of plane but a random in-plane orientation, and grain size determined by the size of the CNO nanosheets. We prepared an LDP with composition of $Bi_3Fe_5O_{12}$ on a CNO-coated GGG(111) substrate (Figure 3, Figure S7), resulting in epitaxial Bi garnet ($Bi_3Fe_5O_{12}$, BiIG) on the exposed areas of GGG between the nanosheets, and Fe-rich $BiFeO_3$ (Fe:BFO) on the CNO. The continuity of the lattice fringes indicates epitaxy at the BiIG/GGG interface, whereas the Fe:BFO region is a single crystal with its $(00l)_{pc}$ lattice fringes parallel to the substrate (Figure 3bc). $BiFeO_3$ typically grows as a rhombohedral structure on STO but a tetragonal phase with $c$ = 0.466 nm is stabilized when the film is compressively strained[37]. In the LDP, both rhombohedral and tetragonal $BiFeO_3$ phases are present (Figure 3d), with the tetragonal phase ($c$ = 0.459 nm) forming at the CNO interface, consistent with compressive strain from the CNO (in-plane lattice parameter 0.386 nm), and the rhombohedral phase developing with film thickness (Figure 3c). The CNO flakes therefore provide an alternative perovskite template, surviving the PLD process to generate LDPs consisting of textured perovskite and single crystal garnet. Furthermore, the CNO sheets can be patterned lithographically prior to growing the LDP (Figure S7ef).

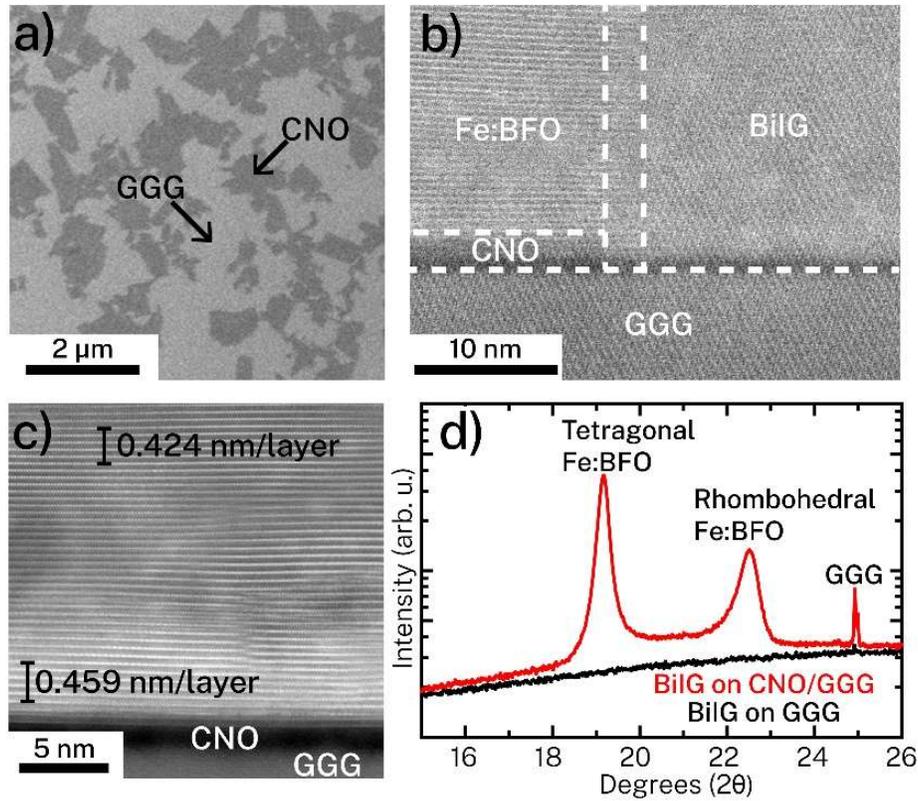

**Figure 3: LDPs synthesized using nanosheet templates.** (**a**) SEM image of a GGG substrate partly covered with CNO nanosheets (dark contrast). (**b**) Cross-sectional STEM image of the Bi$_3$Fe$_5$O$_{12}$ LDP grown on a substrate such as that of (a). (**c**) cross-sectional STEM image of the perovskite region of the LDP at the substrate-film interface, showing the relaxation of out-of-plane lattice parameter with increasing thickness. (**d**) XRD patterns for the Bi$_3$Fe$_5$O$_{12}$ LDP and a BIG film grown on unpatterned GGG in the same run. The LDP shows tetragonal and rhombohedral Fe:BFO peaks which are absent in the film grown on a bare garnet substrate.

**Multiferroic LDPs**

To prepare magnetoelectric LDPs, we take advantage of the ferroelectric behavior of BFO and the ferrimagnetism, low damping, and magnetooptical activity of Bi-substituted YIG across a range of Bi contents, while incorporating a conductive LSMO seed layer to enable the application of an out-of-plane electric field across the thickness of the perovskite phase. A 21 nm LSMO polycrystalline seed grown on GGG exhibited a resistivity of $0.159 \pm 0.35$ Ωcm. To grow the LDP, a Bi$_x$Y$_{3-x}$Fe$_5$O$_{12}$ composition was chosen to produce a stoichiometric garnet and a Fe-rich perovskite, noting that a high Bi concentration favors ferroelectric behavior in the perovskite, while raising the damping in the garnet [38,39]. We selected the composition $x = 2.25$ for the LDP to obtain FM and FE polymorphs, and used (111) $Gd_3Sc_2Ga_3O_{12}$ (GSGG) substrates to provide a close lattice match to the high-Bi-content garnet phase (lattice parameter of 1.256 nm predicted for $Bi_{2.25}Y_{0.75}Fe_5O_{12}$[40] vs. 1.2554 nm for GSGG).

To characterize the FE phase, polycrystalline perovskite of composition $Bi_{2.25}Y_{0.75}Fe_5O_{12}$ (Fe:BYFO, 35.4 nm thick) was grown on a 21 nm thick LSMO seed layer on a GSGG substrate, and Ta/Au contacts were overlaid. A series of positive up negative down (PUND) measurements (Figure 4b, Figure S10) was conducted at various voltages, showing a polarization of 17.01 ± 0.70 µC/cm$^2$ and -11.64 ± 0.77 µC/cm$^2$ for +10 V and -10V respectively. This asymmetry is due to the dissimilar contacts on the top (Ta/Au) and bottom (LSMO) of the ferroelectric. An I-V curve with its corresponding P-E loop (Figure S3b) is in qualitative agreement with the polarization in Figure 4b. Additionally, the out-of-plane polarization was switched at +/-8 V applied to the probe tip in a PFM box-in-box measurement (Figure 4ef). The Fe:BYFO therefore shows ferroelectric properties despite the presence of both Y and excess Fe. These findings are consistent with reports of Fe-rich BFO (with Bi:Fe = 1:1.2) and Y-substituted BFO (BYFO), both of which have been explored separately to reduce the leakage current of BFO[41–43]. BYFO was also shown to have an increased magnetic moment (~2.5 kA/m)[42] compared to BFO[44], while our Fe:BYFO had a net magnetic moment of ~20 kA/m.

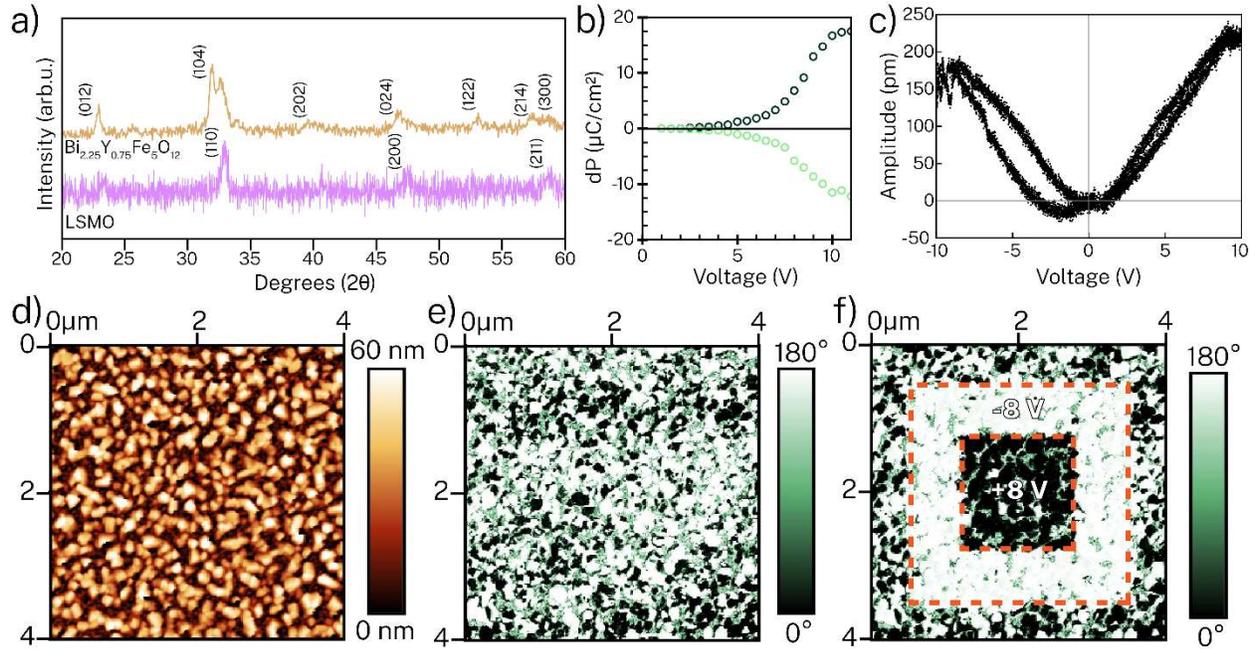

**Figure 4: Ferroelectric phase in the LDP.** (**a**) GIXRD scans of a 21 nm thick LSMO seed layer grown on a GSGG substrate (lower data) with perovskite peaks indexed, and of 35.4 nm of $Bi_{2.25}Y_{0.75}Fe_5O_{12}$ grown on top (upper data) with peaks indexed as the rhombohedral R3c phase. (**b**) Polarization (dP/2) calculated from a series of square-wave PUND measurements ranging from 0 to 11 V, saturating at approximately 10 V. Error bars for the PUND data are smaller than the markers. (**c**) Piezoresponse amplitude of 10 V bipolar sweep revealing butterfly curves. (**d**) AFM topographic scan of the surface of the polycrystalline perovskite film. (**e**) Corresponding PFM phase data showing vertical polarization. (**f**) PFM scan after box-in-box writing at +/- 8V revealing a remnant polarization.

For the FM phase, garnet films of stoichiometry $Bi_{2.25}Y_{0.75}Fe_5O_{12}$ were grown on (111) GSGG and exhibited good structural quality, low strain, and a small negative magnetic anisotropy (anisotropy field $H_{k,eff}$ = -17.7 ± 0.71 mT, $K_u$ = -0.867 kJ/m$^3$ measured with FMR), Figure 5a-d. The low total anisotropy is attributed to competition between growth-induced (magnetotaxial) perpendicular anisotropy[45] and in-plane shape anisotropy. In the range of 3 – 11 GHz, the FMR linewidth is 1.7 – 1.8 mT, comparable to that of other BiYIG compositions[38,46] and about 10 times that of YIG, while the Gilbert damping was 3.5 x 10$^{-4}$. The low film strain promotes low damping, which facilitates magnon transport (spin wave propagation) over long distances.

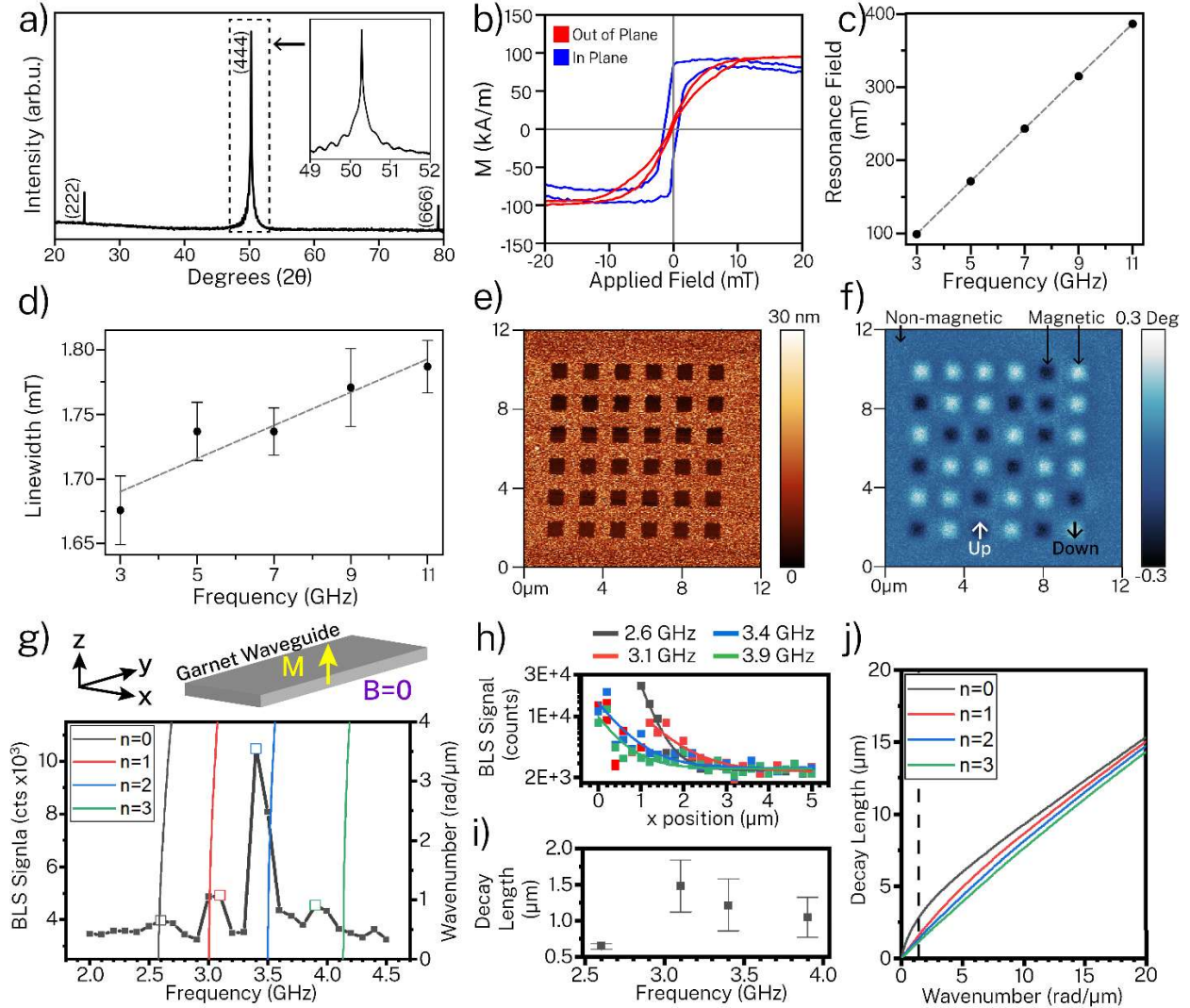

**Figure 5: Ferrimagnetic phase in the LDP.** (**a**) Diffraction peaks of a 42 nm thick film of $Bi_{2.25}Y_{0.75}Fe_5O_{12}$ on GSGG (111). The film peak overlaps the substrate but Laue fringes are visible. (**b**) VSM measurements of the magnetic hysteresis loops of the same film. (**c**) FMR data for resonant frequency vs. field, and (**d**) linewidth vs. frequency of the same film. (**e**) AFM micrograph of 42 nm thick $Bi_{2.25}Y_{0.75}Fe_5O_{12}$ LDP consisting of garnet squares surrounded by a perovskite phase. (**f**) MFM micrograph of the same region, displaying 'up' and 'down' (black and white) remanent magnetization of the garnet squares. (**g**) BLS signal measured approximately 1 μm away from the antenna with the waveguide in a single-domain remanent state in zero magnetic field. The solid lines show the simulated dispersion relation (wavenumber vs. frequency) for modes of order n for the range of wavenumber produced by the antenna. (**h**) BLS intensity dependence on the propagation distance. The experimental data (squares) were fitted with an exponential decay. (**i**) The resulting decay lengths from the data of (f) are on the order of ~1 μm. (**j**) Simulated decay lengths vs k, with the excitation cut-off of the antenna depicted as the black dashed line, leading to observable decay lengths similar to those measured in (i).

The 42 nm thick $Bi_{2.25}Y_{0.75}Fe_5O_{12}$ LDP shown in Figure 5ef includes 800 nm square garnet regions surrounded by a perovskite matrix. From the atomic force microscopy (AFM) topographic image (Figure 5e), the garnet regions are smoother and lower in height by ~9 nm than the perovskite, corresponding to the thickness of the LSMO seed layer (11 nm). The square garnet regions exhibit perpendicular magnetic anisotropy with anisotropy field $H_{k,eff}$ = 196 mT (i.e. $K_u$ = 9.6 kJ/m$^3$) measured by micro-focused Brillouin light scattering (BLS)[47,48] microscopy (Figure S12a). AC-demagnetization of the 800 nm squares led to a distribution of 'up' and 'down' magnetizations showing black and white in the magnetic force micrograph (Figure 5f), characteristic of single domain behavior.

The perpendicular magnetic anisotropy of a larger area of garnet in the LDP, measured by BLS, was $K_u$ = 13.0 ± 0.06 $kJ/m^3$ and exchange stiffness $A_{ex}$ = 4.69 ± 0.01 pJ/m (Methods). The difference in anisotropy compared to the garnet grown on as-received GSGG is attributed to annealing-induced smoothing of the substrate surface during the crystallization of the LSMO seed layer (Supplementary Note 6).

These results show the successful synthesis of an LDP consisting of regions of both ferrimagnetic garnet with perpendicular magnetization, and ferroelectric Fe-rich bismuth yttrium ferrite. We demonstrated the magnonic functionality of the LDP by measuring spin wave propagation within garnet stripes that act as magnon waveguides (Supplementary Note 7). We patterned a 2 μm-wide Au microstrip excitation antenna on one end of a 0.5 μm-wide garnet waveguide, which was then magnetized into a single domain state using an out-of-plane field. Wide-field Kerr microscopy confirmed that the single domain state remains stable at remanence. We swept the excitation frequency from 2.0 to 4.5 GHz and measured the BLS signal at a position 1 μm away from the antenna. The measured BLS spectrum exhibits four peaks which can be attributed to individual waveguide modes, matching approximately the calculated dispersion relation of the waveguide (Figure 5g). Measuring the BLS intensity up to 5 μm along the waveguide, Figure 5hi, gives decay lengths of the individual waveguide modes ranging from 0.5 μm to 2.0 μm. This is consistent with the low group velocities of the magnons excited by the micron-sized antenna[49] within the range from *k*-vectors in the range from 0 to $\frac{\pi}{2\mu m} \approx$ 1.5 rad μm$^{-1}$; a narrower antenna would give higher *k* and longer propagation lengths due to the higher magnon group velocity (Figure 5j).

**Magnetoelectric Coupling in LDPs**

To illustrate the emergent functionality of the LDPs, we demonstrate two examples of magnetoelectric coupling: the voltage-modulation of spin wave propagation and of magnetooptical hysteresis. In each case the magnetic response is interpreted as a result of magnetoelastic anisotropy in the garnet caused by strain-coupling to the piezoresponse of the ferroelectric perovskite at the vertical interfaces.

First, in the $Bi_{2.25}Y_{0.75}Fe_5O_{12}$ LDP we measured changes in magnon propagation in BiYIG waveguides upon application of an electric field to the adjacent Fe:BYFO. The piezoresponse of the Fe:BYFO (Figure 4c) leads to strain transfer into the BiYIG waveguide causing a change in the anisotropy and thus the magnon dispersion relation of the waveguide. Figure 6a shows a schematic of the device, consisting of a 2.5 μm wide, 42 nm thick garnet waveguide flanked with Fe:BYFO. Assuming that the out-of-plane piezoelectric strain is transferred to the garnet over a distance of Λ from the interface, we calculate the resulting anisotropy profile of the waveguide (Figure 6b). We then calculate the lowest frequency magnon modes expected at the center of the waveguide for different values of Λ = 2.5t to 10t, where t is the BiYIG thickness (Supplementary Note 8, Figure 6c).

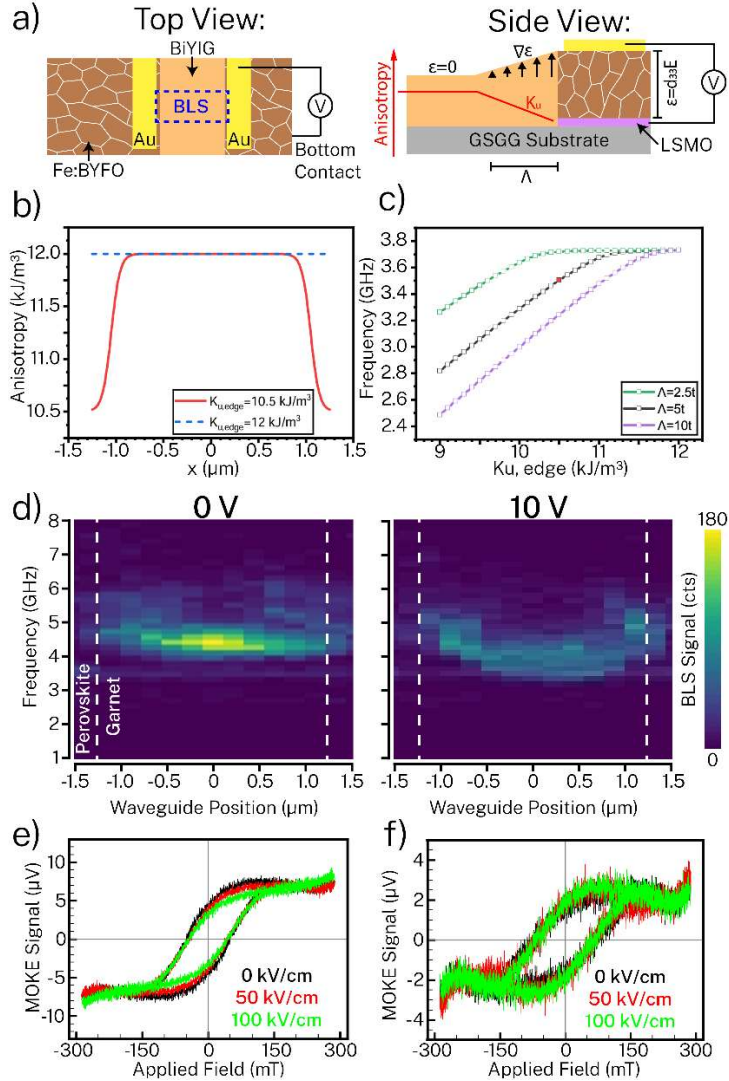

**Figure 6: Magnetoelectric coupling in LSMO-seeded $Bi_{2.25}Y_{0.75}Fe_5O_{12}$ and CNO-seeded $Bi_3Fe_5O_{12}$ laterally differentiated polymorphs.** (a) Illustration of magnetoelectric device consisting of BiYIG waveguide surrounded by Fe-rich BYFO. Top view (left) shows electrodes; side view (right) indicates schematically the piezoelectric strain in the BYFO is transferred into the BiYIG over a distance of $\Lambda$. (**b**) Model for anisotropy near the edges of the 2.5 µm wide BiYIG waveguide, assuming magnetoelastic anisotropy arises from strain transfer from the perovskite phase which decays over a length of $\Lambda = 5t_{Bi:YIG}$ from the interface (red line). The blue line is the unstrained zero-voltage case. (**c**) Simulated center frequency of the magnons in the 2.5 µm waveguide as a function of the change in edge anisotropy, for different values of decay length $\Lambda$ from $2.5t_{Bi:YIG}$ to $10t_{Bi:YIG}$. The red data point corresponds to the conditions in (b). (**d**) Example BLS spectra taken at 0 and 10 V. The center frequency is reduced (by ~250 MHz) when 10 V is applied, as predicted by the model. (**e,f**) Room temperature polar MOKE hysteresis measurements of (**e**) a CNO-seeded $Bi_3Fe_5O_{12}$ LDP showing a reduction in remanence and loop area on applying an electric field, and (**f**) a garnet-only control sample, showing no effects of electric field.

Experimentally, a 250 MHz magnon frequency shift was observed when the voltage applied across the Fe:BYFO was varied between 0 and 10V (Figure S12d). In the model, for $\Lambda = 5t$, this frequency change could be produced by a change of anisotropy of only 1.5 kJ/m$^3$ at the edge of the garnet waveguide. This anisotropy change is well within the magnetoelastic anisotropy of 5.76 kJ/m$^3$ expected from full strain transfer from the Fe:BYFO into the BiYIG, based on the piezoresponse of 200 pm (Figure 4c) and the elastic properties and magnetostriction of the BiYIG. Therefore, strain transfer from the perovskite phase can readily account for the voltage-induced frequency shifts of the 2.5 μm wide BiYIG waveguide (Supplementary Note 8). We further demonstrate voltage-modulation of the modes in a 10 μm-wide waveguide in Figure S13.

A second example of magnetoelectric coupling is shown in the $Bi_3Fe_5O_{12}$ LDP composite grown on CNO-coated GGG (Figure 6f). An electric field of up to ±100 kV/cm was applied in-plane between pairs of electrodes with 10 μm gaps patterned on the composite, and the change in the hysteresis loop of the LDP in the gap was measured by polar MOKE over an area of μm$^2$ in a field of up to 300 mT. A voltage-driven change is found in the minor hysteresis loop (Figure 6e, Figure S8, Table S1) in which the loop area decreased by over 30% on application of 100 kV/cm, and the coercivity decreased from 46.1 mT to 39.5 mT. In comparison, a single-phase garnet film of the same composition showed no significant voltage-induced change in its MOKE loop area (Figure 6f, Table S1).

**Conclusions**

Epitaxial stabilization of the crystal structure of a film grown on a heterogeneous substrate provides extensive opportunities for fabrication of multifunctional polytype composites in which two phases of the same composition have very different properties. We exploit this concept to develop garnet-perovskite composites in which the perovskite regions are ferroelectric, while the garnet regions are ferrimagnetic with low damping, bulk-like magnetization, high magnetooptical activity, and perpendicular magnetic anisotropy. Ferrimagnetic garnets provide exceptional properties including the lowest achievable damping which is essential for magnonic devices; tunable anisotropy; km/s domain wall velocities; magnetization compensation temperatures; angular momentum compensation temperatures yielding antiferromagnet-like dynamics; and magnetooptical activity. Incorporation of garnets into LDPs with a ferroelectric phase not only enables convenient patterning of waveguides and other structures without the need for a garnet etch but also yields multiferroic magnetoelectric composites with demonstrated voltage modulation of magnon dynamics and hysteresis loops.

This patterned-epitaxy method is expected to be extendable to other stoichiometries and phase pairings to produce a variety of multifunctional LDPs. For example, $Lu_xY_{3-x}Fe_5O_{12}$ offers another candidate garnet/perovskite system considering the low Gilbert damping of the garnet and defect-mediated ferroelectricity in the perovskite phase[50]. Interactions at the garnet/perovskite interfaces, including exchange bias and Dzyaloshinskii-Moriya interactions, are amplified in vertical nanocomposites compared to layered films, and voltage modulation may then be expected to control magnetic textures and reversal mechanisms. Voltage modulation of the magnetooptical response of BiYIG 'pixels' surrounded by ferroelectric regions will be useful in creating magneto-optical spatial light modulators enabled by the LDP architecture. We therefore anticipate a range of device applications for composite polymorph films.

**Acknowledgements**

Y.K. acknowledges support from Fujikura Limited.  P.E.L., K.H., and C.A.R. acknowledge support from Fujikura Limited; the National Science Foundation DMR 2323132; and use of the shared facilities of MIT.nano. K. H. was also supported by JSPS KAKENHI grant numbers 20J12063 and 22KJ3113. O.W. and M.U. acknowledge project CZ.02.01.01/00/22_008/0004594 (TERAFIT). CzechNanoLab project LM2023051 funded by MEYS CR is gratefully acknowledged for the financial support of the BLS experiments at CEITEC Nano Research Infrastructure.

# Methods

**Fabrication of LDPs**

Laterally differentiated polymorphs (LDPs) were grown on 10x10 mm (111) $Gd_3Ga_5O_{12}$ (GGG) and (111) $Gd_3Sc_2Ga_3O_{12}$ (GSGG) substrates from MTI Corporation. The as-received substrates were sonicated for 5 minutes in acetone and isopropyl alcohol to remove debris and were then prepared for spin coating by applying an $O_2$ plasma clean at a power of 200 W for 5 minutes in a Glow Research AutoGlow barrel asher. The garnet surface was then treated with an adhesion promoter, SurPass4000, by spin coating at 3000 rpm for 30 s. After spinning, the sample was rinsed with DI water for 30 s and then baked on a hotplate at 180 °C to dry the substrate. Polymeric resist (poly(methylmethacrylate), 950PMMA A2) was spun onto the surface of the substrate at 1500 rpm for 60 seconds before baking on a 180 °C hotplate for 2 min. The resulting resist layer was ~100 nm thick (Figure S1a). To prevent charging during lithography, DisCharge was spun onto the surface of the resist at 1000 rpm for 2 minutes.

An Elionix-F125 electron beam lithography (EBL) system was used to pattern the resist (Figure S1b). When exposing areas with a feature size less than 1 μm, a 2 nA current was used to deliver a dosage of 1200 μC/cm$^2$ to the selected region of the resist. For large feature sizes, greater than 1 μm, a 40 nA current was used instead. For all exposures, proximity effect correction was calculated through GenISys Beamer lithography software in conjunction with GenISys Tracer Monte Carlo simulator.

DisCharge was then removed from the surface with a 30 s D.I. water rinse. The exposed sample was then submerged into a solution of 3:1 IPA:MIBK at -5 °C for 60 s for developing. Immediately after developing, the sample was rinsed with IPA to remove any remaining MIBK and to raise the temperature of the sample and prevent condensation (Figure S1c).

The developed sample was then coated with the seed layer (Figure S1d). Both $SrTiO_3$ (STO) and $La_{0.7}Sr_{0.3}MnO_3$ (LSMO) thin films were prepared via PLD using a Compex pro KrF laser (248 nm wavelength) to ablate a target of the respective composition. The films were deposited with a shot rate of 10 Hz under a base pressure of $1\times10^{-5}$ Torr (i.e. no oxygen was added to the chamber during growth) while the substrates were at 25 °C. These conditions were selected to reduce the conformality of the film to allow for easier liftoff, and to prevent damaging the resist. The laser energy was 300 mJ for the STO depositions (20 nm/1000 shots) whereas the LSMO was grown at an energy of 350 mJ (10 nm/1000 shots). In either case, the as-deposited film was amorphous. Following seed deposition, the remaining resist was removed by sonicating the sample in acetone for 10 minutes. Finally, the seed layer was crystallized via heat treatment in a Riko MLA 5000 rapid thermal annealer at 750 °C for 5 minutes under 1 sccm of $O_2$ (Figure S1e).

An alternative method for creating LDPs incorporated $Ca_2Nb_3O_{10}$ (CNO) as the perovskite seedlayer, where CNO consists of 2D flakes of μm-scale diameter prepared by exfoliation of bulk CNO in solution. (111) GGG double-side polished substrates were hydrophilized by $O_2$ plasma using an asher. Then, a dimethyl sulfoxide ($C_2H_6SO$, DMSO) dispersion containing 0.69 wat% CNO and 0.036 wt% graphene oxide (GO) was spin-coated on the GGG substrates at 1500 rpm for 1 hour to prepare a hybrid monolayer film of CNO and GO (Figure S2ab). The GO was then removed by $O_2$ plasma etch from the hybrid film exposing the GGG underneath. The substrates therefore consist of GGG covered with CNO flakes with coverage determined by the ratio of GO to CNO in the solution (Figure S2c). The CNO coverage was 50-60% according to SEM measurements and ~55% by comparing FR and MOKE measurements of LDPs to those of all-garnet films grown on pristine GGG (Figure 3e).

The LDP layers were prepared using the same PLD system as the seed layers. The films with a stoichiometry of $Y_3Fe_5O_{12}$, $Y_4Fe_4O_{12}$ ($YFeO_3$), $Bi_{2.25}Y_{0.75}Fe_5O_{12}$, and $Bi_3Fe_5O_{12}$ were grown from a single target of the same composition (Figure S1g and S2d). For intermediate compositions, alternating shots of two targets were used. Y-Fe and Bi-Fe oxide films were deposited under a pressure of 10 mTorr $O_2$ (after pumping to $1 \times 10^{-5}$ base pressure) at a shot rate of 10 Hz. Substrates were held at 900 °C for Y-Fe oxide films and 650 °C for Bi-Fe oxide films, 85 mm above the ablated target. Bi-Y-Fe-O films were deposited under a pressure of 50 mTorr $O_2$ ($1 \times 10^{-5}$ base pressure) at a shot rate of 2 Hz. Substrates were held at 900 °C, 85 mm above the ablated target. These temperatures represent the heater setpoint which is approximately 150˚C higher than the actual substrate temperature.

**Electrode Fabrication for ferroelectric testing**

For top electrodes used for ferroelectric testing, samples were prepped for spincoating with a $O_2$ plasma clean at a power of 200 W for 5 minutes in a Glow Research AutoGlow barrel asher. The surface was then treated with SurPass4000, by spin coating at 3000 rpm for 30 s. After spinning, the sample was rinsed with DI water for 30 s and then baked on a hotplate at 180 °C to dry the substrate. Polymeric resist (poly(methylmethacrylate), 950PMMA A4) was then spun on the surface at 1100 RPM for 60 s and then baked at 180 °C on a hotplate for 2 min. The resulting resist layer was ~400 nm thick. DisCharge was then spun onto the surface of the resist at 1000 rpm for 2 minutes. A pattern consisting of squares ranging in side length from 100 μm to 10 μm was exposed in an Elionix-F125 EBL system. A 40 nA current was used to deliver a dosage of 1200 μC/cm$^2$. The same proximity effect correction used in the pattering of LDPs was also used to pattern these contacts. These samples were then developed in 3:1 IPA:MIBK at -5 °C for 300 s for developing. A ~5 nm Ta adhesion layer followed by a ~50 nm Au layer was then deposited via DC magnetron sputtering. Finally, the remaining PMMA was removed via liftoff in a solution of 80 °C N-Methylpyrrolidone (NMP).

**Electrode and excitation antenna fabrication for spin wave excitation**

Multi-step lithography was used for fabrication of the top electrodes and antenna structures for spin wave excitation. First, alignment marks were fabricated using optical lithography. The sample was cleaned in acetone and isopropyl alcohol baths with ultrasound for 3 min consecutively, then dried on a hotplate to remove excess moisture. A positive optical resist AZ MIR 701 was spun onto the sample at 4000 rpm for 60 s with a subsequent bake at 90 °C for 90 s. Alignment marks were patterned using UV laser lithography (ML3 Baby, Quantum Design). Then Ti(4 nm)/Au(20 nm) stack was evaporated onto the sample (e-beam evaporator system, BESTEC) and followed by a lift off. The position of the alignment crosses was measured using an electron microscope (MIRA3, Tescan). The final structures were then prepared by electron beam lithography. After the same cleaning procedure, a positive e-beam resist AR-P 679.04 was spun on the sample at 4000 rpm for 60 s, resulting in a 270 nm thick resist layer, and the sample was baked for 180 s at 150 °C. To prevent charging, the AR-PC 5092 Electra was spun at 2000 rpm for 60 s with a subsequent bake for 120 s at 90 °C. The patterning was performed in a scanning electron microscope (MIRA3, Tescan) equipped with a laser interferometer stage and dedicated pattern generator (Raith LIS). After the exposition, the Electra was dissolved in a DI water bath and the exposed resist was then developed in the AR 600-56 developer for 3 min, stopped by an isopropanol rinse. Finally, the electrode stack of Ti(5 nm)/Au(60 nm) was deposited using the e-beam evaporator followed by a lift-off.

Materials characterization is described in Supplementary Information Section 1.

19. Choi, H. K. *et al.* Hierarchical Templating of a BiFeO3–CoFe2O4 Multiferroic Nanocomposite by a Triblock Terpolymer Film. *ACS Nano* **8**, 9248–9254 (2014).

20. Aimon, N. M., Choi, H. K., Sun, X. Y., Kim, D. H. & Ross, C. A. Templated Self-Assembly of Functional Oxide Nanocomposites. *Advanced Materials* **26**, 3063–3067 (2014).

21. Kumar, A. *et al.* Magnetoelectric Vertically Aligned Nanocomposite of YFeO3 and CoFe2O4. *Advanced Electronic Materials* **8**, 2200036 (2022).

22. Kim, T. C. *et al.* Self-assembled multiferroic epitaxial BiFeO3–CoFe2O4 nanocomposite thin films grown by RF magnetron sputtering. *J. Mater. Chem. C* **6**, 5552–5561 (2018).

23. Xiao, M., Shen, D. & Huang, J. Interface engineering for enhanced memristive devices and neuromorphic computing applications. *International Materials Reviews* **70**, 205–247 (2025).

24. Song, J. & Wang, H. A material design guideline for self-assembled vertically aligned nanocomposite thin films. *J. Phys. Mater.* **8**, 012002 (2025).

25. Qin, H., Both, G.-J., Hämäläinen, S. J., Yao, L. & van Dijken, S. Low-loss YIG-based magnonic crystals with large tunable bandgaps. *Nat Commun* **9**, 5445 (2018).

26. Schmidt, G., Hauser, C., Trempler, P., Paleschke, M. & Papaioannou, E. Th. Ultra Thin Films of Yttrium Iron Garnet with Very Low Damping: A Review. *physica status solidi (b)* **257**, 1900644 (2020).

27. Carothers, K. J., Norwood, R. A. & Pyun, J. High Verdet Constant Materials for Magneto-Optical Faraday Rotation: A Review. *Chem. Mater.* **34**, 2531–2544 (2022).

28. Dong, G. *et al.* Ferroelectric Phase Transition Induced a Large FMR Tuning in Self-Assembled BaTiO3:Y3Fe5O12 Multiferroic Composites. *ACS Appl. Mater. Interfaces* **9**, 30733–30740 (2017).
23

# Supplemental Information:

# 1. Extended Methods and Characterization

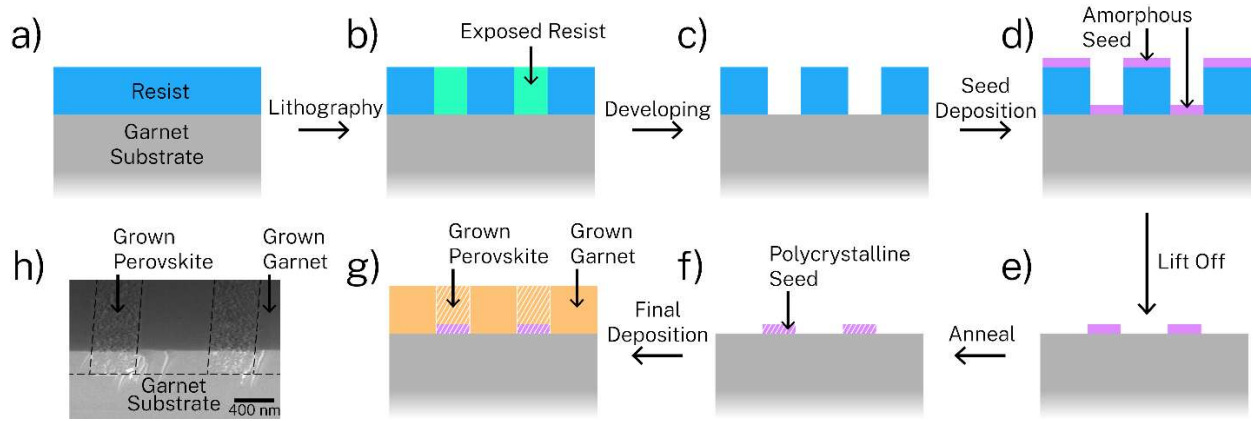

**Figure S1:** Detailed fabrication pathway for the creation of LDPs using a perovskite seed. (**a**) Spin coating of resist onto garnet substrate. (**b**) Selective exposure of resist through EBL. (**c**) Removal of resist regions that were exposed during EBL. (**d**) Seed layer growth via PLD. (**e**) Liftoff to remove remaining resist and regions to reveal regions of garnet substrate. (**f**) Annealing step to crystallize the seed layer within an RTA to achieve a garnet-perovskite two phase substrate. (**g**) Final deposition step of oxide via PLD, to form garnet-perovskite derived LDPs. (**h**) SEM micrograph of cross-sectional view of 220 nm thick stoichiometric $Y_3Fe_5O_{12}$ LDPs grown on (111) GGG substrate, seeded with 25 nm of STO. The sample is tilted, and the darker contrast in the top half of the image corresponds to the top surface while the lighter contrast is the cross-section.

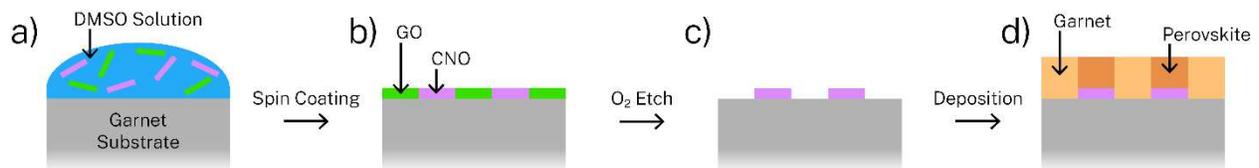

**Figure S2:** Preparation of CNO seed layers. (**a**) Flakes of CNO and GO are spin coated from DMSO solvent onto a garnet substrate. (**b**) A monolayer of CNO/GO remains on the substrate. c) Oxygen etching removes the GO leaving the CNO partly covering the surface. d) Growth of the LDP by PLD.

**Structural, magnetic, and ferroelectric characterization**

*Scanning transmission electron microscopy (STEM) and Scanning Electron Microscopy (SEM):* A Raith Velion FIB-SEM was used to create the cross-sectional lamellae analyzed in a Thermo Fisher Scientific Themis Z G3 Cs-Corrected S/TEM and a Super X detector for energy dispersive xray spectroscopy (EDS). To enhance visibility of lighter elements during imaging, integrated differential phase contrast was used[1]. Additionally, for the high magnification STEM imaging of individual phases, drift compensation frame integration in Velox was used. A Zeiss Gemini 450 SEM was used to collect all SEM micrographs. Prior to investigation, samples were coated with 5 nm of AuPd with a PELCO SC7 Sputter Coater to prevent sample charging.

*X-ray Diffraction (XRD):* High-resolution X-ray diffraction (HRXRD) scans of single crystalline garnet samples were collected using a Rigaku Smartlab diffractometer with a Ge-(220) double-bounce monochromator with a Cu X-ray source. Grazing incidence X-ray diffraction (GIXRD) scans of polycrystalline films were conducted on the same equipment.

*Vibrating Sample Magnetometry (VSM):* The in-plane and out-of-plane magnetic hysteresis loops were collected using a DMS 880A VSM. This instrument was also used to demagnetize samples prior to magnetic force microscopy (MFM); a 500 mT field was applied to the out of plane direction and was alternated in direction and decreased in intensity by a factor of 0.95 until the field approached 0 mT.

*Atomic Force Microscopy (AFM):* A Bruker Dimension Icon SPM was used to perform all AFM, MFM, and PFM measurements. Bruker TESPA-V2 cantilevers operating in tapping mode were used to collect all AFM scans. MFM measurements used Bruker MESP-V2 tips with a lift mode. Lift height ranged from 75-150 nm. For PFM scans, samples underwent additional preparation; PELCO High Performance Silver Paste was applied to exposed LSMO and connected to a metal disc to ground the conductive seed layer. PFM measurements used a Bruker SCM-PIT-V2 cantilever operating under a contact mode. For amplitude vs voltage curves, a sweep frequency of 0.4 Hz was used.

*Polar Magneto-Optic Kerr Effect (MOKE) measurement:* Polar MOKE measurements were performed using light with a wavelength of 660 nm and a spot size of ~8 um. The light was linearly polarized and directed onto a sample at near-normal incidence, and the reflected light was measured with an analyzer using a Si photodetector as a function of magnetic field perpendicular to the sample. To measure the effect of electric field, electrodes were patterned with a spacing of 10 μm on the surface of the sample, and an in-plane voltage of up to 100 V was applied between them while the MOKE loop was measured from the region between the electrodes. A linear background from the 0 V loop was fitted and subtracted from the loops measured at different voltages.

*Ferroelectric testing:* A Precision Premier II tester was used to measure the ferroelectric response with PUND and P-E hysteresis loop measurements. For PUND measurements, a series of square-wave pulses ranging from 0.5 V to 11 V were applied with a pulse width of 1 ms, and a delay of 10 ms. After an initial, unmeasured, negative pulse, a positive pulse was applied and the total (switching plus non-switching) polarization ($P^*$) was measured at the end of the pulse. Then a second positive pulse was applied to measure the non-switching polarization ($P^\wedge$). Two negative pulses were likewise sequentially applied to acquire $P^*$ and $P^\wedge$ in the negative direction. The remanent-only polarization (dP/2) was finally calculated from the difference between these two polarizations, the values of which are plotted in main text Figure 4b.

$$dP = P^* - P^\wedge$$

Subsequently, P-E hysteresis loops were constructed from a PUND-like series of applied voltages using triangle waveform pulses, a protocol used successfully to characterize leaky $(Hf,Zr)O_2$[2]. The advantage of using this method compared to square-wave PUND is that it allows the voltage at which ferroelectric switching occurs to be captured, and thus the P-E loop can be obtained from a series of measurements. For example, in an 11 V measurement, a -11 V triangle pulse was applied to the contacts to ensure a correct initial polarization. Then, two identical pulses with a magnitude of 11 V were applied, the first of which captured the ferroelectric switching response plus the dielectric and resistive components, whereas the second pulse just captures the dielectric and resistive components. To extract the ferroelectric response, the I-V curve of the second pulse was subtracted from the first pulse. This same process was carried out in the negative direction with two more pulses with a magnitude of -11 V. These two monopolar loops were then stitched together to get a full I-V hysteresis curve of the isolated ferroelectric response. Finally, these results were integrated to calculate the polarization (P), where *A* is the contact surface area, *I* is the current, and *t* is time:

$$P = \frac{1}{A}\int I\,dt$$

The waveform and corresponding *I* vs *t* response of one such measurement is displayed in Figure S3a. Figure S3b shows the P-E loop and the current calculated from analysing a series of data like Figure S3a measured at varying voltages.

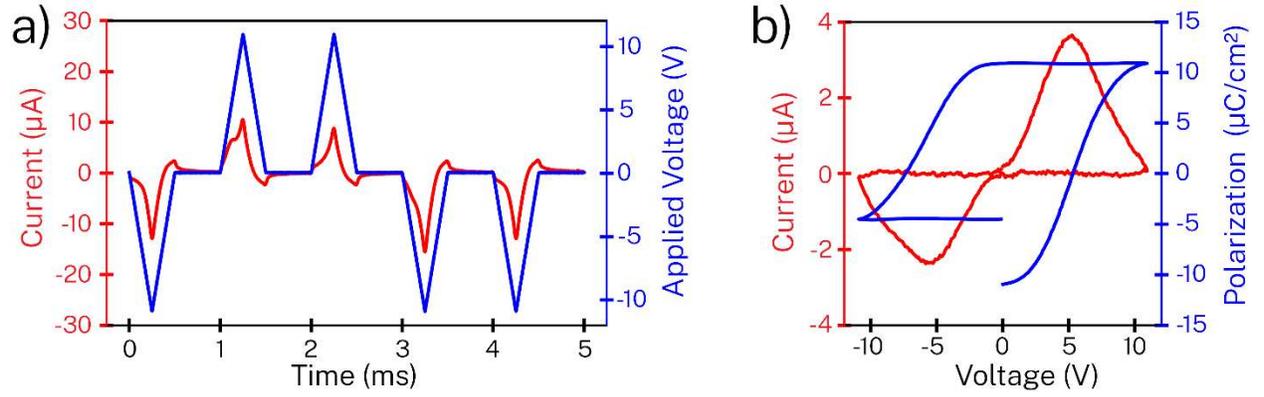

**Figure S3:** (**a**) Example P-E loop measured with triangular voltage waveforms and (**b**) the calculated ferroelectric polarization vs. voltage and current vs. voltage curves.

*Ferromagnetic Resonance (FMR):* FMR was conducted with a custom built broadband FMR system to determine the damping and the net anisotropy of unpatterned garnet thin films. An in-plane magnetic field was swept from μ₀H = 0 to 400 mT, while an RF magnetic field was passed through a coplanar waveguide ranging in frequency from 3 to 11 GHz. The resulting scans were fitted to the sum of symmetric and anti-symmetric Lorentzian functions to extract linewidth ($\Delta H$) and resonance field ($H_{res}$), where S, A and B are fitting variables:

$$\frac{dP}{dH} = A\frac{4(H-H_{res})\Delta H}{[4(H-H_{res})^2 + (\Delta H)^2]^2} + S\frac{(\Delta H)^2 - 4(H-H_{res})^2}{[4(H-H_{res})^2 + (\Delta H)^2]^2} + B$$

The effective anisotropy field ($H_{k,eff}$) of the film was then determined by fitting the resonance field to the applied frequency ($f$) with the Kittel equation:

$$f = \frac{\mu_0 \gamma}{2\pi} \sqrt{H_{res}(H_{res} - H_{k,eff})}$$

Finally, the Gilbert damping ($\alpha$) and the inhomogeneous broadening ($H_{ih}$) were fitted from:

$$\Delta H = H_{ih} + \frac{\alpha}{\gamma} f$$

*Microfocused Brillouin Scattering Light Spectroscopy (uBLS):* A custom-developed optical setup was used to measure Brillouin light scattering spectra[3]. A single-mode laser Cobolt Samba (532 nm) was used as the light source. The spectral purity of the laser light was improved by a Fabry-Perot filter (TCF-2, from Table Stable). The incident power on the sample was 3 mW. An optical microscope with active stabilization was used to compensate mechanical drift of the sample (THATec Innovation). The light was focused and collected through the same objective (Zeiss LD EC Epiplan-Neofluar 100 × /0.75 BD). The inelastic frequency shift was measured with a tandem Fabry-Perot interferometer (FP-2HC interferometer, Table Stable)[4]. To generate the magnetic field, we used a water-cooled GMW 5403 electromagnet powered by two KEPCO BOP20-20DL power supplies and a predefined current field calibration at the sample position.

To fit spin wave resonances, we used a semi-analytical model. The first step was to find the equilibrium angle of the magnetization $\theta$ with respect to the normal of the film (i.e. 0 degrees is out of plane, and 90 degrees is in plane), by minimizing the total energy:

$$\frac{E_{tot}}{V(\Theta)} = \frac{E_{ani}}{V} + \frac{E_{demag}}{V} + \frac{E_{Zeeman}}{V} = -K_1 \sin^2(\theta) - \frac{\mu_0}{2} M_s^2 \cos^2(\theta) - B_{ext} M_s \cos(\theta)$$

The obtained angle was then used to calculate the resonant frequency for $k = 0$:

$$f^2 = \left(\frac{\gamma}{2\pi}\right)^2 \left(B_{ext} \cos\theta + B_c + \left(\frac{2K_1}{M_s} - \mu_0 M_s\right) \cos^2\left(\frac{\pi}{2} - \theta\right) + \frac{2A_{ex}}{M_s}\left(\frac{n}{d}\right)^2\right)$$
$$\cdot \left(B_{ext}\cos\theta + B_c + \left(\frac{2K_1}{M_s} - \mu_0 M_s\right) \cos\left(2\left(\frac{\pi}{2} - \theta\right)\right) + \frac{2A_{ex}}{M_s}\left(\frac{n}{d}\right)^2\right)$$

The fit was performed simultaneously on the fundamental and first perpendicular standing spin wave mode and the exchange constant ($A_{ex}$), uniaxial anisotropy in the out-of-plane direction ($K_1$) and field independent of the magnetization direction ($B_c$) were varied in the fit. The saturation magnetization was fixed at $M_s = 98$ kA/m and thickness $d = 42$ nm. The origin of $B_c$ can be attributed either to the lattice coupling or (higher order) cubic anisotropy.

*Simulation of the dispersion relation and decay lengths:* The dispersion relation was calculated using simulation software TetraX[5,6]. The waveguide cross-section was modeled using a finite-element method and translation symmetry along the long edge of the waveguide was included. The characteristic length of each element in the out-of-plane direction was 5 nm and in the in-plane direction was 5 nm (25 nm) for a 500 nm (2.5 µm)-wide waveguide. The static magnetization configuration was calculated using energy minimization and dispersion was obtained by an eigenfrequency solver. The linewidth ($\delta f$) was calculated from the mode profiles[7] and inhomogeneous broadening of 50 MHz was considered, based on measurements with the omega-shaped antenna on another BiYIG layer. With this assumption the lifetime is about 19.3 ns and has little dependence on a wavenumber. The group velocity was obtained as the derivative of the dispersion relation $\left(v_g = \frac{d\omega}{dk}\right)$ and was used together with the lifetime ($\tau = \frac{1}{\delta f}$) to calculate decay lengths ($\delta = \tau v_g$).

## 2. Seed layer and LDP examples

Figure S1 shows the full sample fabrication process. The growth of the seed layers of LDPs is critical to the creation of multiphase structures. Figure S4a shows an example of a seed layer of STO deposited over a resist pattern consisting of 800 nm wide squares with 100 nm gaps and subsequently lifted off. Figure S4b,c shows the same 220 nm thick $Y_3Fe_5O_{12}$ LDP as in Figure 1e,f consisting of a series of garnet hexagons with a side length of 770 nm, spaced 400 nm apart on a (111) GGG substrate. The topography of the 10 nm STO seed layer is shown in Figure 1e.

Figure S4d,e shows SEM and AFM images of a seedlayer pattern with arbitrary shapes including curves with a minimum feature size of 50 and 100 nm. These fine structures were maintained after the LDP growth of a 165 nm thick $Y_{3.5}Fe_{4.5}O_{12}$ film (Figure S4f).

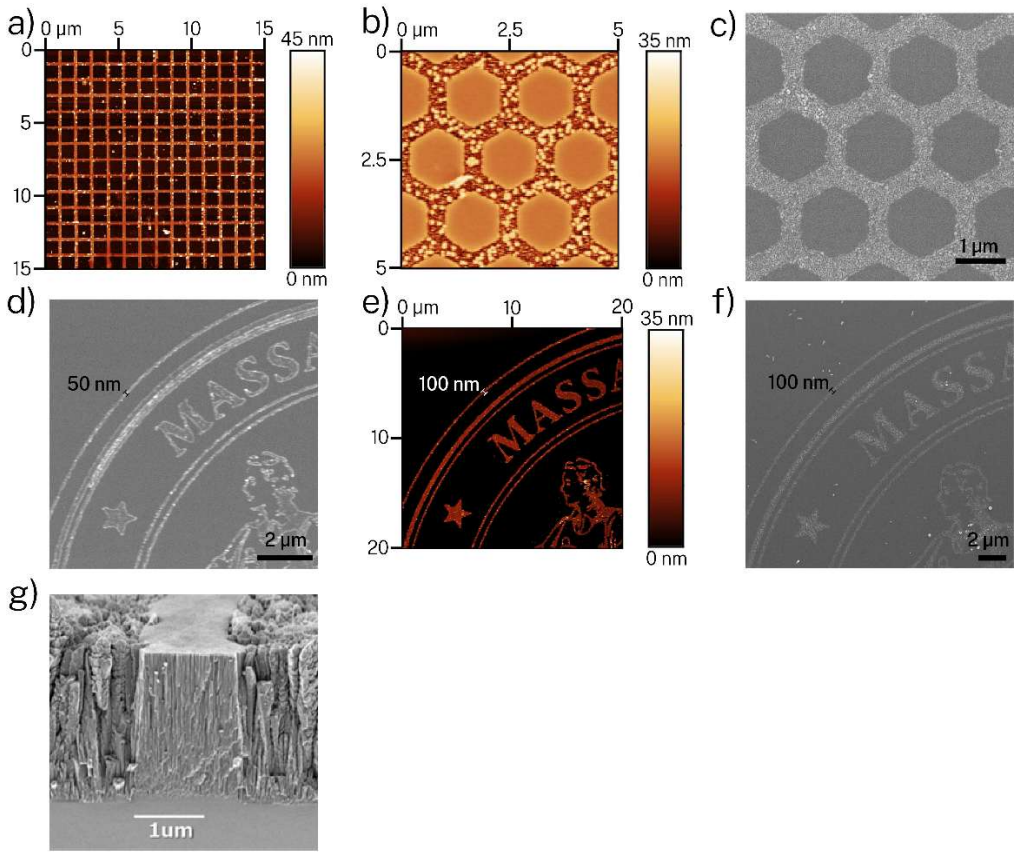

**Figure S4:** (**a**) 16 nm thick STO patterned into a square array with 100 nm wide lines. (**b,c**) 220 nm thick $Y_3Fe_5O_{12}$ LDPs patterned into a hexagonal array by deposition on a 10 nm thick STO seed layer on a (111) GGG substrate analyzed with AFM (**b**) and SEM (**c**). (**d**) SEM image of 12 nm thick STO patterned into the Massachusetts Institute of Technology seal with 50 nm minimum

feature size. (**e**) AFM of 12 nm thick STO patterned into the Massachusetts Institute of Technology seal, with 100 nm minimum feature size. (**f**) SEM of the patterned Massachusetts Institute of Technology seal in (e) with 100 nm minimum feature sizes, after depositing 165 nm of $Y_{3.5}Fe_{4.5}O_{12}$ forming a LDP. (**g**) A LDP with thickness 2.2 µm with Y:Fe = 3:5, showing persistence of the two phases despite the nonideal stoichiometry of the perovskite.

### 3. Phase formation and film stoichiometry

A range of $Y_{3+X}Fe_{5-X}O_{12}$ compositions (where X ranges from 0 to 1) were deposited on (111) GGG substrates to explore their structural and magnetic properties. A 382 nm thick film with a 3:5 Y:Fe cation ratio (stoichiometric YIG) grew as expected as a garnet evident by the (444) diffraction peak present in the HRXRD scans in Figure S5, along with a bulk-like room temperature saturation magnetization of 135 kA/m. A 165 nm thick film with a Y:Fe cation ratio of 3.5:4.5 on (111) GGG formed a garnet structure with an increased out-of-plane lattice parameter. The excess Y is likely occupying octahedral garnet sites to form $Y_{Fe}$ antisite defects which increase the unit cell volume. The reduction in superexchange coupling lowers the Curie temperature, and thus the room temperature magnetization[8], as evidenced by a reduced saturation magnetization of 126 kA/m. Similarly, the 350 nm thick 4:4 Y:Fe cation ratio film was still stabilized as a garnet phase, with a higher out-of-plane lattice parameter and a decrease in saturation magnetization to 43 kA/m.

The formation of a polycrystalline perovskite-derived phase on top of a STO seed layer is shown in Figure S5c. First, 12 nm of STO was deposited onto a (111) GGG substrate at $10^{-5}$ torr at 25 °C and was crystallized via RTA as described in the Methods. The resulting film was analyzed via GIXRD which indicated the formation of a polycrystalline STO. Then, a 382 nm thick film with composition $Y_3Fe_5O_{12}$ (stoichiometric YIG) was deposited onto the seeded substrate at 900 °C. The film showed peaks corresponding to an orthoferrite despite its garnet stoichiometry.

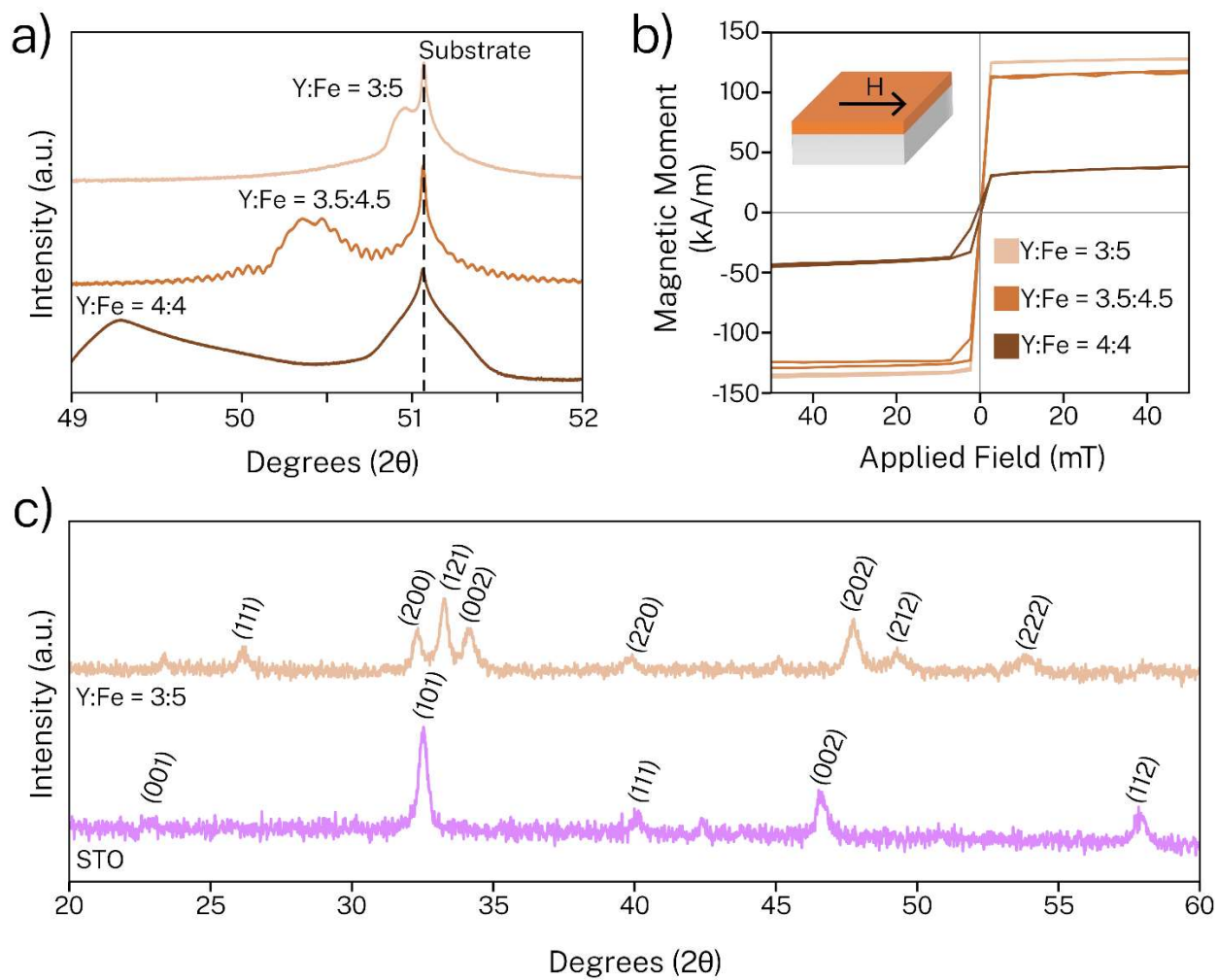

**Figure S5:** (**a**) HRXRD scans of Y:Fe oxides of varying ratios and thicknesses grown on a (111) GGG substrate. For all scans the (444) substrate GGG peak is present as well as (444) film peaks from the garnet phase film. (**b**) VSM scans of the films shown in (a) with in-plane field, all of which show an in plane easy axis. Increased excess of Y led to smaller magnetic moment. (**c**) GIXRD scans of 12 nm STO seed layer with peaks indexed to the cubic perovskite (lower data, purple), and a 382 nm film of composition $Y_3Fe_5O_{12}$ (upper data, tan) grown on top of the seed layer, with peaks indexed to the orthorhombic orthoferrite phase. The STO film was grown directly on a (111) GGG substrate at room temperature, and was crystallized at 750 °C for 5 minutes.

In addition to the cross-sectional stoichiometric analysis conducted with STEM (Figure 2b), SEM-EDS line scans were conducted across the surface of the sample to confirm a constant stoichiometry. Figure S6 shows an EDS line scan of a 220 nm thick $Y_3Fe_5O_{12}$ LDP film with a 1 µm wide seeded perovskite region separated with 1 µm of garnet. Across the observed area, Y, Fe, and O are present in the same relative ratio (Figure S6b), with an average stoichiometry of $Y_{3.04}Fe_{5.29}O_{11.67}$. As expected, this composition is within the margin of error of EDS of $Y_3Fe_5O_{12}$. Prior to EDS analysis, this film was sputtered with 3 nm of AuPd to prevent charging, and as a result both metal peaks are present within the EDS spectra in Figure S6c.

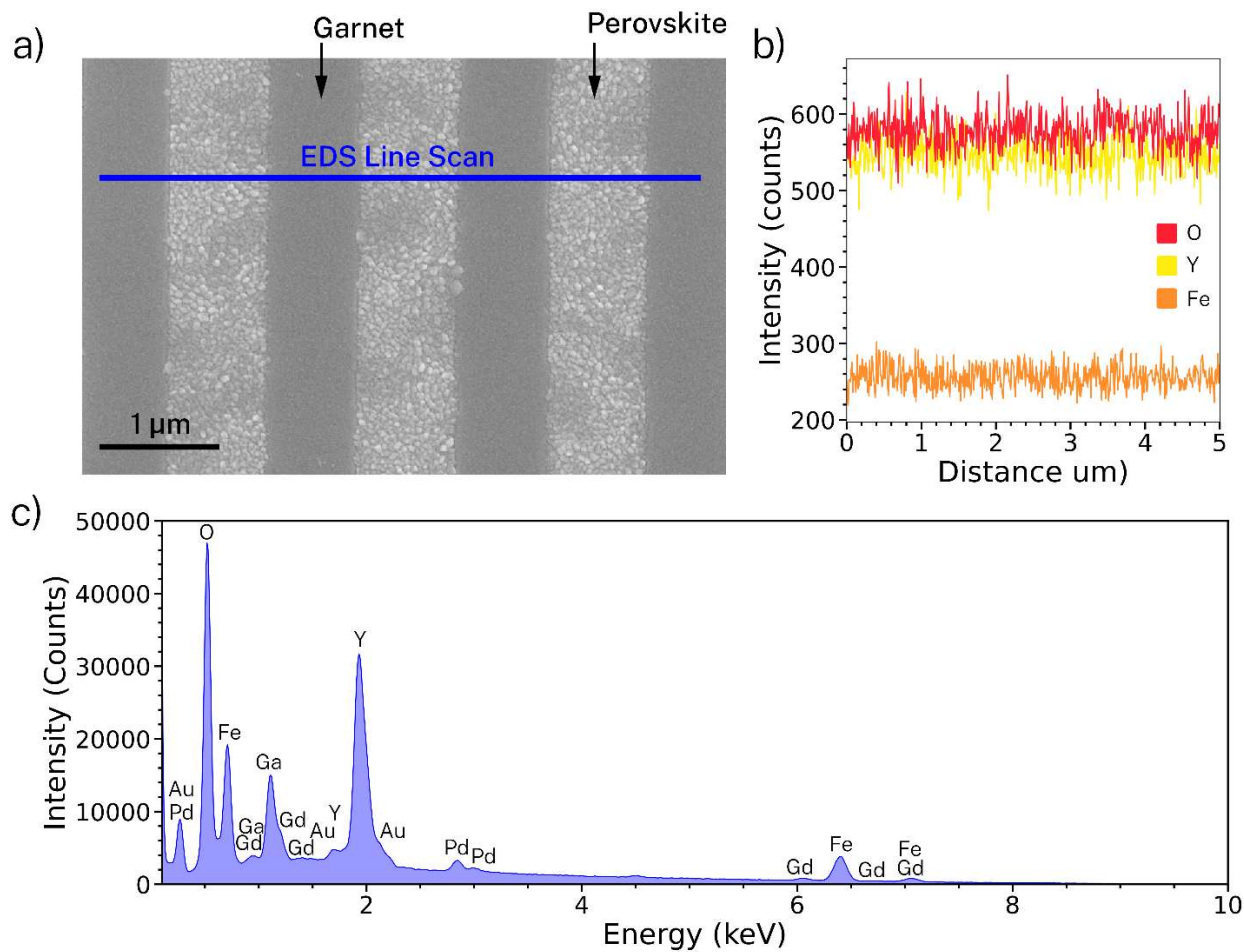

**Figure S6:** SEM/EDS analysis of 220 nm thick $Y_3Fe_5O_{12}$ LDP thin film with a STO seed layer. (**a**) SEM image of the location in which an EDS line scan was collected. (**b**) EDS line scan displaying constant film composition (Y,Fe,O) with respect to lateral position. (**c**) EDS spectra of combined line scan.

## 4. CNO Seed Layer

Figure S2 shows the sample fabrication process. Figure S7a,b show AFM images of a hybrid film of CNO and GO before and after $O_2$ plasma etching, respectively (corresponding to the steps in Figure S2b,c respectively). The unetched film consists of approximately one layer of CNO and GO sheets on the substrate with some regions of overlapping sheets. The etch process removes the GO leaving spaces between CNO sheets where the substrate is exposed. Figure S7c,d show AFM and SEM images of the LDP with composition $Bi_3Fe_5O_{12}$ grown on the GO sheets. Clear morphological differences between the garnet and perovskite phases are present. The garnet (BiIG) phase forms regions with fine scale topography on a length scale of ~100 nm. The CNO sheets template (001)-oriented perovskite single crystals, unlike the STO seed, but the top surface of the perovskite regions is less smooth and has larger topographical features than the garnet phase.

To demonstrate lithographic patterning of a CNO seed layer, we made a resist pattern on a GGG substrate with 5 µm wide trenches. CNO was coated over the resist and attached to the substrate only within the trenches. Liftoff left the substrate with corresponding regions of CNO. A $Bi_3Fe_5O_{12}$ LDP was then fabricated on the patterned CNO sheets (Figure S7e,f) to produce a garnet film with 5 µm wide perovskite lines. The coverage of the CNO region for this demonstration is about 80%, so some garnet areas are observed inside the 5 µm wide perovskite region (Figure S7f), but by increasing the number of CNO coatings, the coverage can be increased to 100% to produce a textured polycrystalline perovskite-only region.

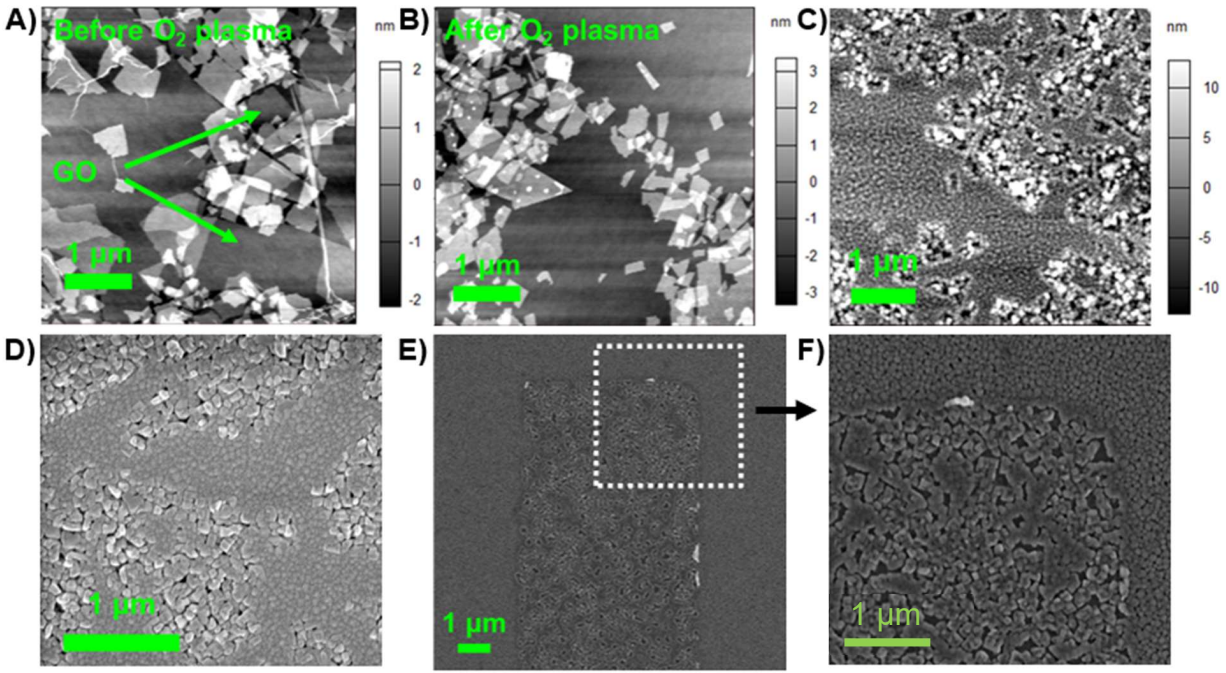

**Figure S7:** (**a,b**) AFM images for the hybrid film of CNO and GO before (a) and after (b) $O_2$ plasma etching. (**c,d**) AFM (c) and SEM (d) images of the $Bi_3Fe_5O_{12}$ LDP. (**e,f**) SEM images of the $Bi_3Fe_5O_{12}$ LDP with a 5 μm wide perovskite line, (**f**) is an enlarged view.

Magnetoelectric behavior of the CNO-seeded films was measured by MOKE, Figure S8 and Figure 6 in the main text. For the garnet/perovskite composite, Figure S8a shows data for positive V and Figure S8b shows data for negative V. Comparing Figure S8a,b shows that reversing the direction of the in-plane electric field yields the same result, as expected for the symmetrical in-plane electrodes. Similar results were obtained for several other measurements. In contrast, for the garnet-only control sample, the MOKE loops showed little or no effect of voltage up to 100 V (Figure S8c). This result is clearly differentiable from the response of the garnet/perovskite composite, where changes in loop shape occur as the voltage is applied. Figure S8 therefore confirms that the garnet/perovskite composite shows a magnetoelectric effect whereas garnet alone shows little or no effect. The magnetoelectric effect is reversible (i.e. the zero-volt loops are the same before and after the voltage is applied).

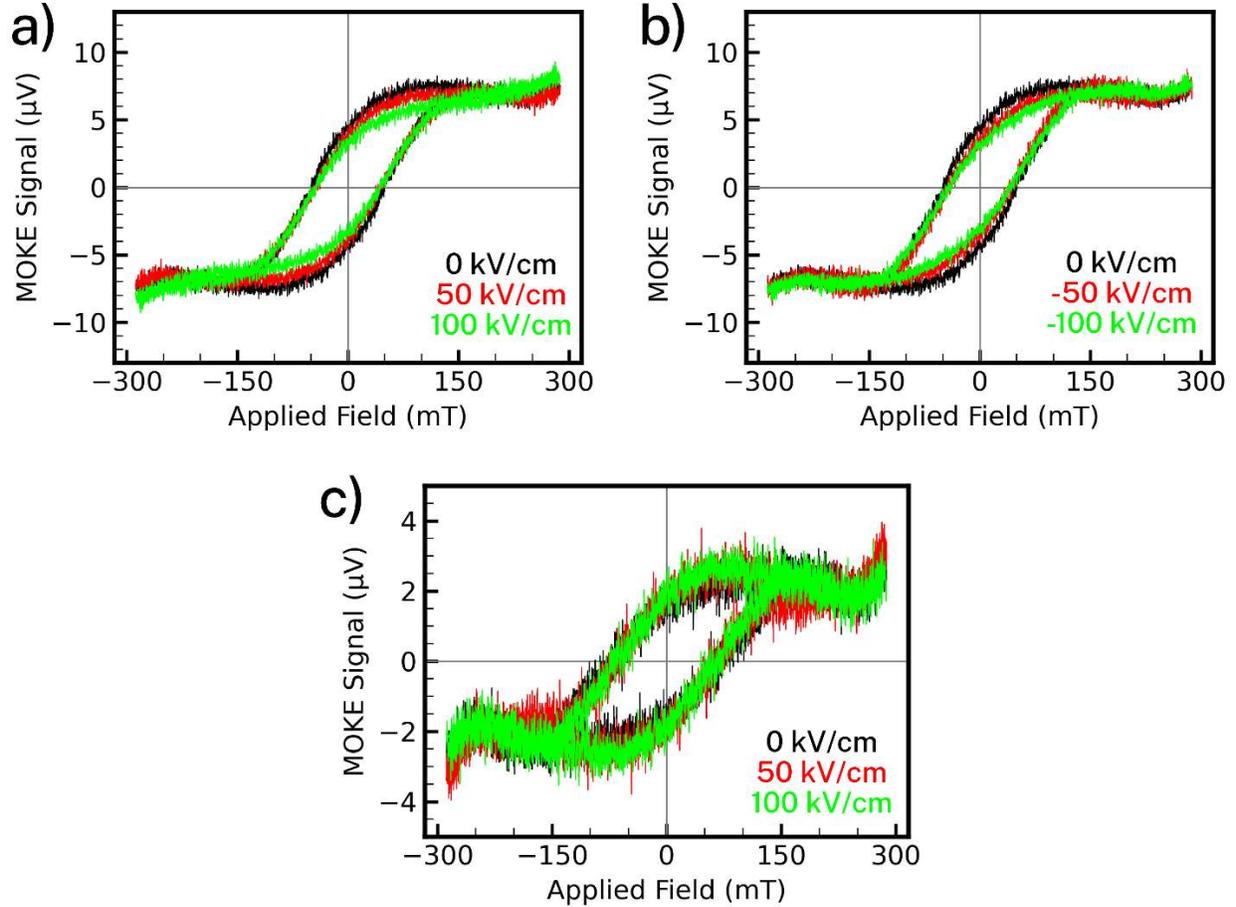

**Figure S8:** Comparison of polar MOKE hysteresis loops with (**a**) a positive applied voltage and (**b**) a negative applied voltage for a CNO seeded BiIG-Fe:BFO LDP. (**c**) Polar MOKE hysteresis loops of single phase BiIG under varying electric electric fields. (The signal for the control sample is noisier due to greater scattering from the gold electrodes.)

To quantify the effect of voltage on the samples, we integrated the loop areas, obtaining the results shown in Table S1. The hysteresis loop area represents the energy dissipated per cycle in reversing the magnetization, and decreases with voltage for the garnet/perovskite composite by an amount (up to 33%) which is much larger than the measurement error (~6%). In contrast, changes in the garnet-only sample loop area (up to 10%) are within the measurement error (13%). This provides unambiguous evidence for a magnetoelectric effect in the composite and not in the control sample.

**Table S1**. Change in MOKE hysteresis loop area with voltage, normalized to the zero-voltage loop area.

| Voltage V | Composite garnet/ perovskite, positive V | | Composite garnet/ perovskite, negative V | | Garnet only sample | |
|---|---|---|---|---|---|---|
| | Loop area | Error +/- | Loop area | Error +/- | Loop area | Error +/- |
| 0 | 1.000 | 0.059 | 1.000 | 0.059 | 1.000 | 0.132 |
| 50 | 0.871 | 0.058 | 0.706 | 0.058 | 1.087 | 0.133 |
| 100 | 0.676 | 0.051 | 0.667 | 0.056 | 1.101 | 0.129 |

The magnitude of the magnetoelectric effect in the MOKE data of the CNO-seeded samples (composition $Bi_3Fe_5O_{12}$) is not directly comparable to the magnetoelectric changes seen in the magnon frequency in the lithographically patterned garnet waveguides (composition $Bi_{2.25}Y_{0.25}Fe_5O_{12}$). First, the electric field applied in the MOKE experiment is considerably lower than that in the waveguide experiment, and is applied in-plane instead of out-of-plane. The field in the MOKE experiment is up to 100V across the electrode spacing of ~10μm, corresponding to an in-plane field of only 100 kV/cm. In the waveguide experiment, 10V is applied across the film thickness of 42nm corresponding to 2.4 MV/cm. The lower field in the MOKE experiment is likely to produce less piezoresponse compared to the waveguide experiment, leading to less strain transfer to the garnet phase and less change in anisotropy. Second, the feature sizes in the composite seeded by CNO are on the length scale of the CNO sheets, order several μm, whereas the features in the waveguide sample are 0.5 μm and above. Since the magnetoelectric effect is a result of strain transfer at the vertical interfaces between the perovskite and garnet, the magnetoelectric effects will be greater for narrower garnet features.

## 5. Fe-rich BYFO (Fe:BYFO)

Further exploration of the electric field response of the $Bi_{2.25}Y_{0.75}Fe_5O_{12}$ thin films grown on LSMO seed layers was conducted via PFM (Figure S9). A small area (1.5 x 1.5 µm) was scanned and reveals that the majority of the surface is ferroelectric, as evidenced by the amplitude micrograph (Figure S9b) containing only a few black regions of zero signal indicating no response, and the phase micrograph (Figure S9c) which contains only a few regions of noise (green, i.e. not 0° or 180° phase) that indicate no out of plane ferroelectric response. The amplitude data of the PFM corresponding to the phase data displayed in Figure 4e,f are displayed in Figure S10g,h, and similarly indicate a surface primarily comprised of ferroelectric grains with non-zero amplitude.

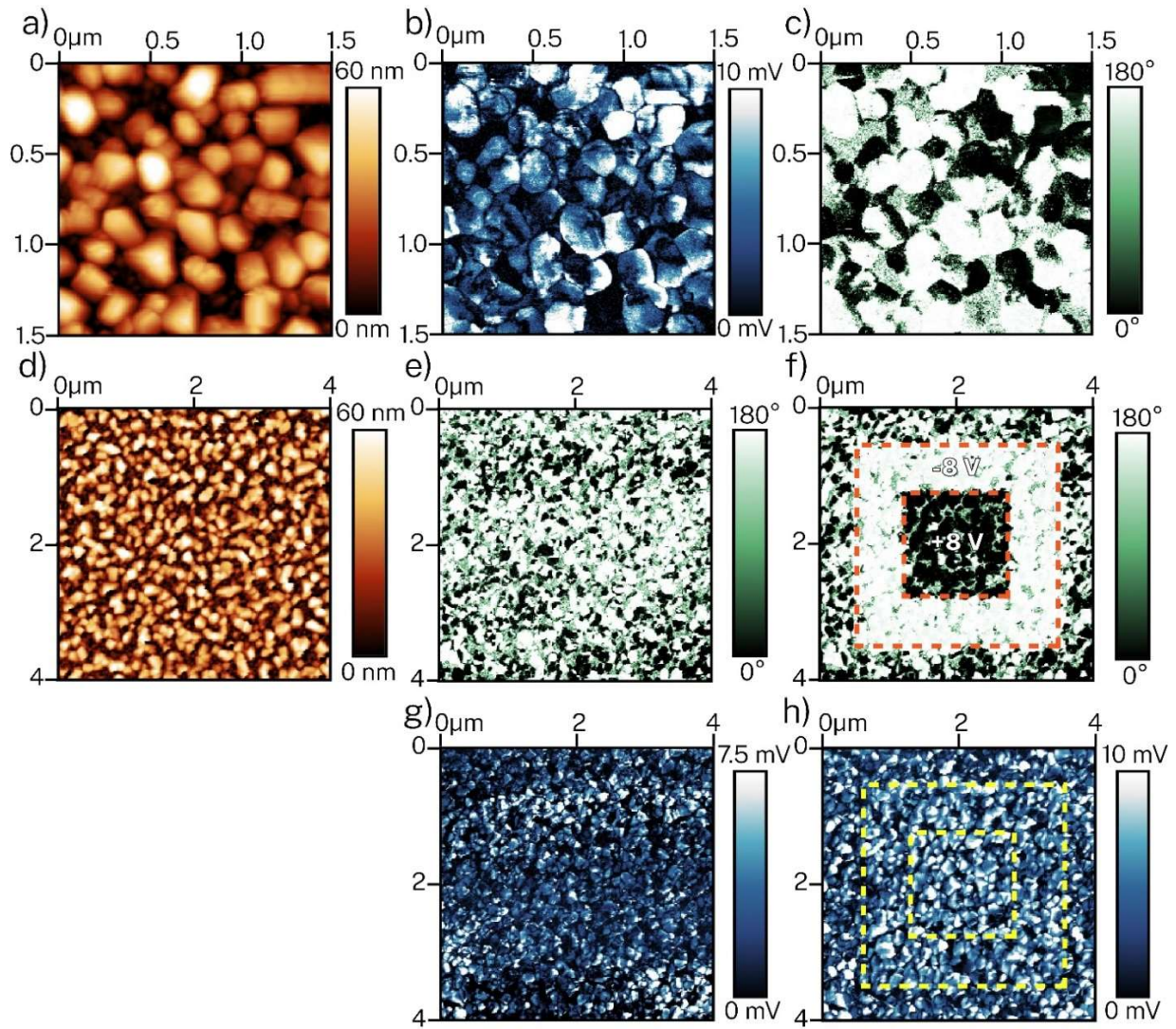

**Figure S9:** PFM scans collected on $Bi_{2.25}Y_{0.75}Fe_5O_{12}$ perovskite film seeded with LSMO on a GSGG substrate. (**a,b,c**) Small area scan of the measured (a) height, (b) amplitude, and (c) phase. (**d,e,f**) Reproduction of AFM and PFM scans displayed in Figure 4d,e,f with corresponding amplitude taken before (**g**) and after poling (**h**).

Figure S10 shows examples of PUND data of a $Bi_{2.25}Y_{0.75}Fe_5O_{12}$ perovskite film for voltages of 5 V and 10 V with square-waveform, showing the method for separating switching and nonswitching polarization. The saturation polarization from the PUND measurement (Figure 4b and Figure S10), 18 µC/cm$^2$, is larger than that from the PUND-like triangular waveform measurements (Figure S3), 11 µC/cm$^2$. This variation is attributed to the different waveforms applied to the ferroelectric: in the case of the PUND measurements, a square wave was applied with a pulse width of 1 ms and a pulse delay of 10 ms. In the case of the PUND-like measurements, a 0.5 ms pulse width and 0.5 ms pulse delay was used.

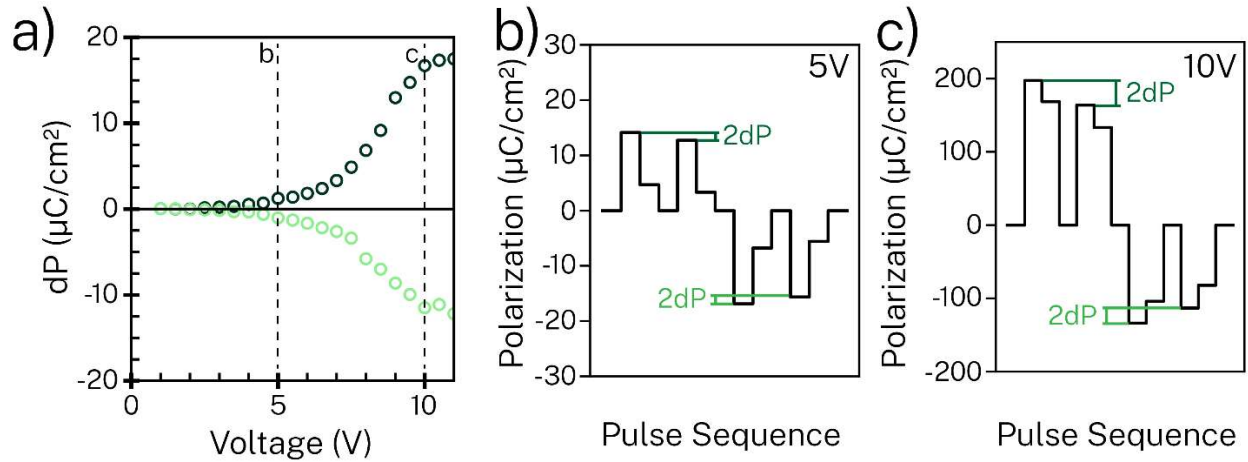

**Figure S10:** (**a**) Compilation of PUND of $Bi_{2.25}Y_{0.75}Fe_5O_{12}$ perovskite film seeded with LSMO, and with a gold top contact, measured across the film thickness (reproduced from Figure 4). Individual PUND measurements used to calculate dP, taken at (**b**) 5 V and (**c**) 10 V pulses superposed on (a).

## 6. Magnetic Anisotropy and Damping of $Bi_{2.25}Y_{0.75}Fe_5O_{12}$ Garnets

Bi can be substituted in YIG across the range of compositions, $Bi_xY_{3-x}Fe_5O_{12}$ although the material becomes unstable for the highest Bi contents and can only be formed as an epitaxial film. The room temperature saturation magnetization varies from 140 kA/m at $x = 0$ to 120 kA/m at $x = 3$, and the Gilbert damping rises from $6.15 \times 10^{-5}$ at $x = 0$ to $1.4 - 2.3 \times 10^{-2}$ at $x = 3$ [9].

The LDP in Figure 5 was made from $Bi_{2.25}Y_{0.75}Fe_5O_{12}$ films grown on (111) GSGG substrates. The net magnetic anisotropy ($K_u$) of the films was calculated from the effective anisotropy field determined using FMR and µBLS

$$K_u = \frac{H_{k,eff} M_s}{2}$$

The magnetic anisotropy of $Bi_{2.25}Y_{0.75}Fe_5O_{12}$ films depends on the patterning and annealing processes that the substrate had undergone. Even though the STO or LSMO seed layer only covered a part of the substrate, the annealing process needed to crystallize it affects the GGG surface and hence the properties of the $Bi_{2.25}Y_{0.75}Fe_5O_{12}$ film.

Comparing two 42 nm $Bi_{2.25}Y_{0.75}Fe_5O_{12}$ films grown on (111) GSGG substrates, the garnet film grown on an unpatterned, untreated substrate (Figure 5a-d) had a net magnetic anisotropy of -0.97 kJ/m³, i.e. in-plane easy axis, whereas the garnet film grown on the substrate partly covered by a LSMO seed layer had a net magnetic anisotropy of +13.0 kJ/m³ (Figure S12a). Here positive $K_u$ indicates an out-of-plane easy axis. The anisotropy includes contributions from magnetoeleastic ($K_{ME}$), magnetostatic ($K_{MS}$), magnetocrystalline ($K_{MC}$), and magnetotaxial or growth induced ($K_{MT}$) anisotropies, according to [10]:

$$K_u = K_{ME} + K_{MS} + K_{MC} + K_{MT}$$

$$K_u = \frac{9}{4}\lambda_{111}c_{44}\left(\frac{\pi}{2} - \beta\right) - \left(\frac{\mu_0}{2}\right)M_s^2 + \frac{K_1}{12} + K_{MT}$$

Because of the close lattice matching between the GSGG substrate and the $Bi_{2.25}Y_{0.75}Fe_5O_{12}$ films (<0.1% mismatch), and the negligible contribution of magnetocrystalline anisotropy, the net magnetic anisotropy is dominated by magnetostatic and magnetotaxial anisotropies. The magnetostatic contribution is -7.60 kJ/m³ by approximating the film as an infinite sheet. The magnetotaxial contributions are then +6.63 kJ/m³ and +20.6 kJ/m³ for the film grown on the unpatterned and patterned substrates respectively.

The difference in the growth induced (magnetotaxial) anisotropy of the film grown on the unprocessed (111) GSGG substrate and the patterned (111) GSGG substrate is attributed to the changes in surface roughness or other structural changes caused during the annealing of the LMSO seed layer. The as-received substrates had root mean square (RMS) roughness of 427 pm (±59 pm) prior to heat treatment, whereas the garnet regions of the patterned substrate, after heat treatment, had RMS roughness of 159 pm (±36 pm). The lower roughness of the patterned garnet likely affected adatom kinetics during growth, resulting in longer lifetimes, and hence increased the observed magnetotaxial anisotropy.

Despite the high concentration of Bi within these films, the Gilbert damping calculated freom FMR was within the same order of magnitude as the damping of YIG films of similar thicknesses grown on GGG. Figure S11 displays the FMR spectra and the Gilbert damping of $Y_{3-x}Bi_xFe_5O_{12}$ thin films on varying substrates. The damping tends to increase with Bi due to an increase in spin orbit coupling that creates new relaxation channels, and a greater amount of inhomogeneous lattice strain from the Bi distribution. The BiYIG synthesized in this work had a lower damping than prior reports.

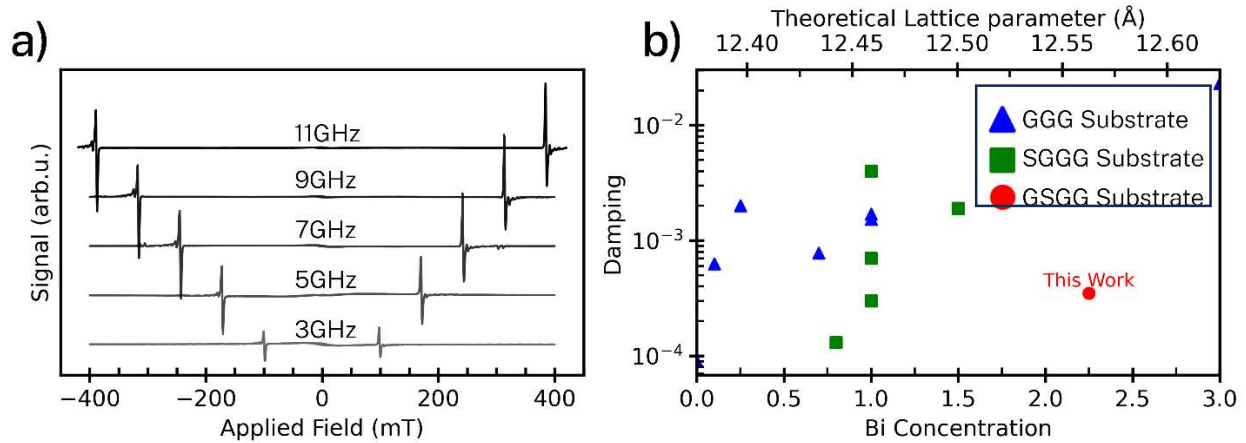

**Figure S11:** (**a**) FMR measurements of an epitaxial $Bi_{2.25}Y_{0.75}Fe_5O_{12}$ thin film for frequencies ranging from 3-11 GHz. (**b**) Comparison of Gilbert damping values of $Y_{3-x}Bi_xFe_5O_{12}$ films from [11–20] grown on a variety of substrates.

# 7. Micro-Brillouin Light Scattering Analysis

Thermal BLS spectra were acquired on 0.5 and 2.5 µm wide LDP garnet waveguides surrounded by perovskite (Figure S12b). In the case of the 2.5 µm wide waveguide, there are two clear spin wave bands at approximately 6.6 GHz and 18 GHz corresponding to the fundamental and first order perpendicular standing spin wave modes, respectively. There is qualitative agreement with the simulation but the frequency of the first perpendicular standing spin-wave mode is lower than the prediction. The differences may be a result of pinning at the waveguide interface. At the edges of the 2.5 µm waveguide, we can also clearly see a shift in the spin wave frequencies caused by the appearance of edge modes which arise due to geometric confinement from the discontinuity in the magnetization.

Surprisingly, there are no spin wave modes quantized across the width of the 0.5 µm waveguide, as one would expect for an ideal narrow waveguide. In these waveguides, width modes should fill the frequency space between 7 and 18 GHz (Figure S12c) but are not observed in the experimental data. This can be explained by the roughness of the waveguide boundaries which disrupts the coherency of the spin wave width modes. The suppression of such spin wave modes could allow for single mode operation of narrow waveguides. We also note that fitting these waveguide spectra yields an anisotropy of $K_\mathrm{u} = 8.8 \pm 0.10\ k\mathrm{J/m^3}$ and that the frequency of the perpendicular standing spin-wave (PSSW) mode is slightly mismatched, which (together with the increased value of $K_\mathrm{u}$ of the 800 nm garnet squares) suggests that anisotropy and exchange coupling are modified in the patterned region.

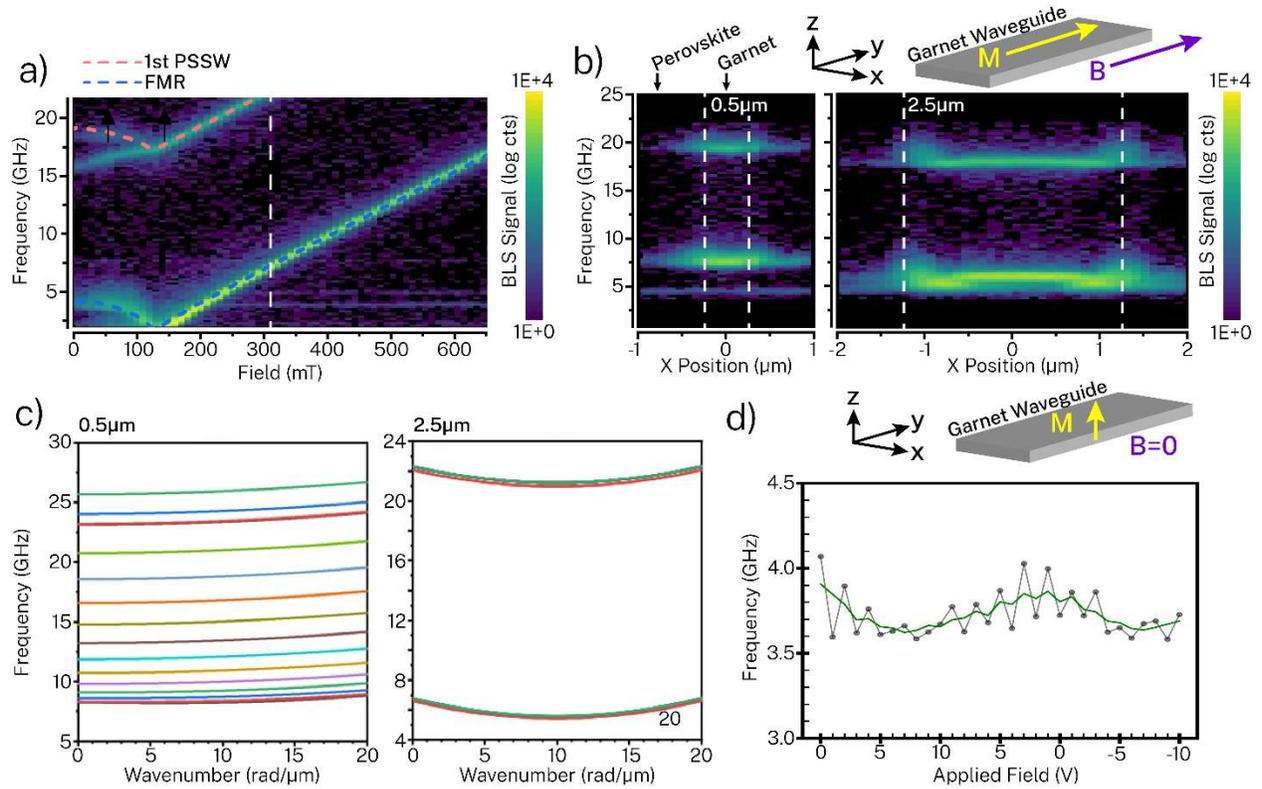

**Figure S12:** (**a**) BLS data measured on an unpatterned region of garnet film in a varying external magnetic field. The dashed lines show the result of the fit, see Methods. The vertical white dashed line marks the magnitude of the external magnetic field applied in panel b. (**b**) BLS spectra scanned across 0.5 μm wide (left) and 2.5 μm wide (right) garnet waveguides embedded in a perovskite matrix with a 307 mT applied field along the long axis of the waveguide. (**c**) Simulated dispersion relation of $Bi_{2.25}Y_{0.75}Fe_5O_{12}$ thin films patterned into waveguides of 0.5 and 2.5 μm in width. (**d**) Effect of applying a 0 → 10 → 0 → -10 V voltage sweep on the center frequency of a 2.5 μm wide waveguide. A cyclic modulation of the frequency by 250 MHz is observed, with a superposed fluctuation attributed to an instrumental artifact.

Additional magnetoelectric coupling measurements were conducted on 10 μm wide garnet $Bi_{2.25}Y_{0.75}Fe_5O_{12}$ waveguides surrounded with ferroelectric perovskite. This was a different device patterned on the same sample as the one measured in Figure 6de. A μBLS spatial scan across the waveguide width was performed to measure the intensity of the spin wave frequencies (Figure S13). After forming the single domain state at 0 V, the voltage was increased to 10 V (2400 kV/cm). Under no external magnetic or electric field, the μBLS spatial scan across the waveguide width displayed three regions of higher frequency in the detected spin-wave frequency with maximum frequencies at approximate positions -3 µm, 0 µm, and +3 µm across the waveguide width (Figure S13). These antinodes suggest a mode superposition across the width of the waveguide. When the voltage was increased from 0 to 6 V, the modulation disappeared, and the detected spin-wave frequency was constant across the waveguide width. However, when voltage was increased to 10 V, the modulation returns. Sweeping back to 0 V, and further down to -10 V, a similar disappearance, and reappearance of this three-fold modulation is observed. This behavior suggests the applied voltage modifies the boundary conditions (surface anisotropy) at the perovskite/garnet interface, and results in a complex, but detectable magnetoelectric effect, where the lateral magnon mode observed at 0 V can be suppressed at certain voltages.

Magnetoelectric coupling measurements in the 2.5 µm wide waveguide in Figure 6d showed an oscillation in the frequency data superposed on the 250 MHz frequency change that occurred when the voltage was cycled. The origin of this oscillation is not known but we believe it is an instrumental artifact because it was not seen in other data such as that of the 10 µm wide waveguide in Figure S15.

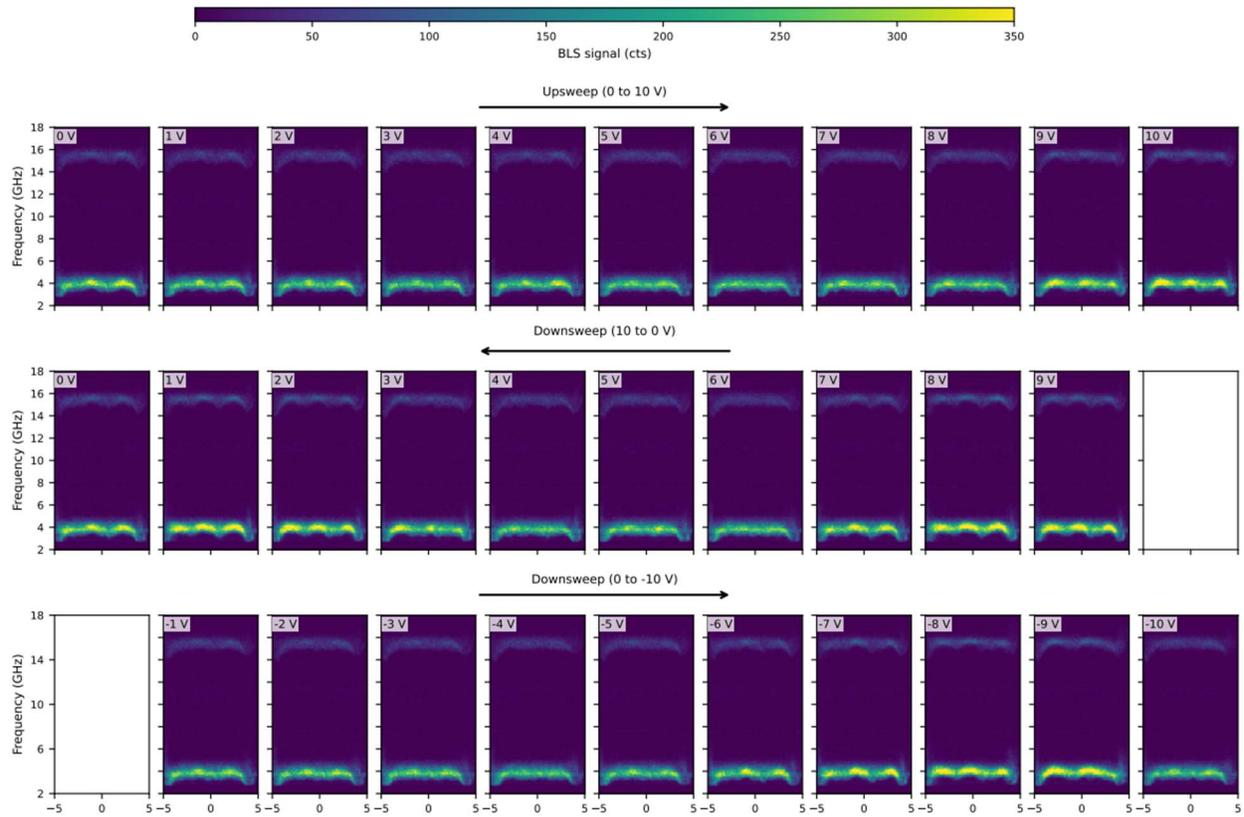

**Figure S13:** BLS spectra of 10 μm wide $Bi_{2.25}Y_{0.75}Fe_5O_{12}$ garnet region while a +/- 10 V sweep is applied to electrodes on the neighboring ferroelectric perovskite.

In the experimentally observed decay of the μBLS data (Figure 6g, red data points), the exponential fitting was only done after a trend was visible at about 1.5 μm from the antenna. The non-monotonic behavior seen close to the antenna may be due to partial blocking of the beam spot by the antenna, a non-linear response of the detector due to the high amplitude of the spin waves, or direct excitation of the magnetization.

# 8 Origin of magnetoelectric coupling leading to BLS frequency change

We briefly summarize the observed phenomenon of a change in BLS frequency with an applied voltage: When a voltage is applied the ferroelectric material, it undergoes a piezoelectric strain. The magnetoelectric response is due to strain transfer between the ferroelectric region and the neighboring magnetic garnet region, resulting in a change in the magnetoelastic anisotropy, and thus the observed magnon frequency. The transferred strain and hence the net anisotropy is not uniform across the garnet waveguide but instead is greatest at the interface between the ferroelectric and magnetic phases.

The out of plane displacement of the sample was directly measured with PFM to be ~200 pm at 10 V (Figure 4c) corresponding to a strain of $\varepsilon = 0.2/42$ or ~0.48%. The voltage-displacement loop itself has a butterfly-like shape expected from a ferroelectric material. Assuming complete strain transfer from the perovskite to the garnet at the vertical interface, we estimate a change in the magnetoelastic anisotropy equal to:

$$K_{ME} = -3/2\, \lambda \sigma$$

where the magnetostriction is $\lambda = -(1 + 0.23z) \times 2.73\, 10^{-6}$ for $Bi_{2.25}Y_{0.75}Fe_5O_{12}$ (with Bi content z = 2.25), i.e. $\lambda = -4\, 10^{-6}$ [26] and the stress $\sigma = E\varepsilon$, with Young's modulus E = 200 GPa for YIG. This yields $K_{me}$ = 5.76 kJ/m³. The strain will decay away from the interface, leading to a non-uniform strain profile and anisotropy across the width of the waveguide which affects the magnon frequency.

To relate the non-uniform strain profile to the magnon frequency change within the 2.5 μm waveguide, we modeled the non-uniform anisotropy (Figure 6b) as a a logistic (sigmoid) function:

$$K_u(x) = K_{u,edge} - \frac{K_{u,edge} - K_{u,film}}{1 + \exp\left[-\frac{4.39}{\Lambda}\left(\frac{W}{2} - |x| - \Lambda\right)\right]},$$

where $K_{u,edge}$ and $K_{u,film}$ is the value of anisotropy at the edge and center of the waveguide respectively. $W$ is the width of the waveguide, and $\Lambda$ is the penetration depth over which the strain imposed by the perovskite, relaxes across the garnet waveguide.

We used values of penetration depth of $2.5 t_{Bi:YIG}$, $5 t_{Bi:YIG}$ and $10 t_{Bi:YIG}$ where t is the thickness of the film. To analyse the effects of strain transfer (due to applied voltage), we calculated the frequencies of the lowest spin wave mode for the different values of edge anisotropies and penetration depth (Figure 6c)[5]. To reproduce the experimentally measured shift of 250 MHz, the model requires a reduction of the edge anisotropy by only 1.5 kJ/m$^3$ when assuming a strain penetration of $5t_{Bi:YIG}$. This is well within the anisotropy change of 5.76 kJ/m$^3$ which is an upper estimate that assumes complete piezostrain transfer from the perovskite to the waveguide. From Figure 6c, as the depth of strain relaxation increases, the reduction of the edge anisotropy required to reproduce the magnon frequency shift of 250 MHz is reduced. Even if the strain is relaxed over just $2.5t_{Bi:YIG}$, a 250 MHz frequency shift is accomplished by an edge anisotropy change of 2.5 kJ/m$^3$. Therefore, strain transfer from the perovskite phase appears well able to explain the voltage-induced frequency shift in the waveguide.

    We attribute the voltage-induced shift in magnon frequency to strain transfer, as opposed to magnon mode coupling between the garnet and the perovskite due to the expected frequency mismatch between magnons in the two phases. The Fe:BYFO antiferromagnon is expected to be within the THz regime, preventing interaction between the GHz ferromagnons observed in the garnet[21,228]. This is further confirmed by the absence of observable magnon modes in the Fe:BYFO within the BLS measurement range. This strain-transfer mechanism is also in qualitative agreement with the trend of reduced energy loss per cycle observed in the MOKE hysteresis loops, Table S1.

**Supplementary References**